\documentclass[preprint,12pt,authoryear]{elsarticle}

\usepackage{multirow}
\usepackage{subcaption}
\usepackage{xspace}
\usepackage{colortbl}
\usepackage[inline]{enumitem}
\usepackage{xcolor, soul}
\usepackage{arydshln}
\usepackage{setspace}
\usepackage{array}
\usepackage{makecell}
\usepackage{graphicx}
\usepackage{listings} 
\usepackage{hanging}
\usepackage{quoting}
\usepackage{amssymb}
\usepackage{xurl}
\usepackage[hidelinks]{hyperref}

  {\list{}{\leftmargin=0.1in\rightmargin=0.0in}  \item[]  }%
  {\endlist}

\definecolor{promptinputcolor}{rgb}{0.055,0.722,0.745}
\definecolor{mygray}{rgb}{0.5,0.5,0.5}
\definecolor{vscode-pink}{RGB}{175,0,219} 
\definecolor{vscode-cyan}{RGB}{38,127,153} 
\definecolor{vscode-darkyellow}{RGB}{121,94,38} 
\definecolor{vscode-darkred}{RGB}{163,21,21} 
\definecolor{vscode-darkblue}{RGB}{0,15,128} 
\definecolor{vscode-green}{RGB}{0,153,0} 
\definecolor{suggestion}{RGB}{229, 88, 219} 
\definecolor{plan}{RGB}{87, 230, 97} 

\begin{document}

\begin{frontmatter}



\title{Reflection-Satisfaction Tradeoff: Investigating Impact of Reflection on Student Engagement with AI-Generated Programming Hints} 


\author[umich]{Heeryung Choi}
\ead{heeryung@umich.edu}

\author[mpisws]{Tung Phung}
\ead{mphung@mpi-sws.org}

\author[umich]{Mengyan Wu}
\ead{mengyanw@umich.edu}

\author[mpisws]{Adish Singla}
\ead{adishs@mpi-sws.org}

\author[umich]{Christopher Brooks}
\ead{brooksch@umich.edu}


\affiliation[umich]{%
    organization={University of Michigan},
    city={Ann Arbor},
    country={USA}
}
\affiliation[mpisws]{%
    organization={Max Planck Institute for Software Systems},
    city={Saarbrücken},
    country={Germany}
}






\begin{abstract}
Generative AI tools, such as AI-generated hints, are increasingly integrated into programming education to offer timely, personalized support.
However, little is known about how to effectively leverage these hints while ensuring autonomous and meaningful learning. One promising approach involves pairing AI-generated hints with reflection prompts, asking students to review and analyze their learning, when they request hints. This study investigates the interplay between AI-generated hints and different designs of reflection prompts in an online introductory programming course. 
We conducted a two-trial field experiment. In Trial 1, students were randomly assigned to receive prompts either \emph{before} or \emph{after} receiving hints, or \emph{no prompt} at all. Each prompt also targeted one of three SRL phases: \emph{planning}, \emph{monitoring}, and \emph{evaluation}. In Trial 2, we examined two types of prompt guidance: \emph{directed} (offering more explicit and structured guidance) and \emph{open} (offering more general and less constrained guidance). 
Findings show that students in the \emph{before}-hint (RQ1), \emph{planning} (RQ2), and \emph{directed} (RQ3) prompt groups produced higher-quality reflections but reported lower satisfaction with AI-generated hints than those in other conditions. Immediate performance did not differ across conditions. This negative relationship between reflection quality and hint satisfaction aligns with previous work on student mental effort and satisfaction. Our results highlight the need to reconsider how AI models are trained and evaluated for education, as prioritizing user satisfaction can undermine deeper learning.
\end{abstract}



\end{frontmatter}



\section{Introduction} \label{sec:intro}

\sloppy{Generative AI systems, such as ChatGPT~\citep{ChatGPT}, are rapidly transforming programming courses. These tools are increasingly employed to enhance educational experiences by offering on-demand instant assistance tailored to students' needs~\citep{DBLP:journals/corr/abs-2402-01580,DBLP:conf/icer/PhungPCGKMSS22,DBLP:conf/ace/Finnie-AnsleyDB22}. Among the various AI-powered educational applications, AI-generated hints have drawn particular attention. AI-generated hints are designed to guide students through complex tasks by providing timely, context-sensitive feedback, which may significantly affect students' problem-solving and help-seeking. Recent studies have indicated teachers' and students' positive reception to using this tool in educational contexts~\citep{DBLP:conf/iticse/PankiewiczB24,burner2025we}.}

However, research has found mixed outcomes regarding the educational benefits of integrating generative AI and AI-generated hints. Some studies reported a negligible or negative effect of students' AI use on their performance, such as a decrease in deep learning~\citep{zamfirescu202561a, bastani2024generative}. In another work by~\citet{DBLP:journals/corr/abs-2307-00150}, the positive impact of AI-generated hints on students’ performance did not last once these hints were removed.

Several studies attributed the discouraging impact of AI on education to \emph{metacognitive laziness}, the tendency of students to avoid and disengage from effortful learning processes~\citep{fan2025beware}. Previous research has associated greater confidence in AI performance and more frequent AI use to increased cognitive offloading, reduced self-reported cognitive effort, and lower metacognitive engagement~\citep{gerlich2025ai, lee2025impact, zhan2025students}. This concern is further underscored by the recent Anthropic Education Report, which showed that approximately 70\% of student-AI interaction was to delegate higher-order thinking to AI agents~\citep{handa2025education}. These findings highlight the pressing need for further research into instructional design and guidance of integrating AI-generated hints to address the issue of metacognitive laziness and more effectively leverage the educational benefits of AI~\citep{DBLP:conf/icer/PhungPCGKMSS22,DBLP:conf/iticse/0001MS0L24,DBLP:journals/corr/abs-2402-01580}.

One promising solution could be reflection-focused interventions. Reflective practices encourage students to critically analyze their thinking, problem-solving strategies, and errors, thereby fostering self-awareness and iterative improvement~\citep{zimmerman2002becoming, tankelevitch2024metacognitive, DBLP:journals/iahe/ChoiJPBJW23}.
These benefits of reflective practices have been well studied and are frequently used as a scaffold to encourage SRL and critical thinking across educational contexts, including environments that provide students with hints. Previous research has demonstrated that prompting reflection is specifically effective in addressing students' metacognitive laziness during their interactions with predefined, less personalized hints ~\citep{DBLP:journals/iahe/ChoiJPBJW23, gerlich2025ai}. When students were prompted to reflect, they seemed to more actively and meaningfully engage with hints; they showed higher self-efficacy, and better immediate and delayed performances~\citep{DBLP:journals/iahe/ChoiJPBJW23, bardach2021power}. Compared to simply preventing unproductive learning behaviors, reflection is often a more effective approach because it not only limits undesirable hint use but also fosters active engagement with learning that is likely to develop into long-term learning behaviors~\citep{aleven2006toward, munoz2013inferring}.

Given these positive impacts of reflection practices, prompting for reflection seemed to be a promising approach to address metacognitive laziness and enhance students' productive engagements with the AI-generated hints. However, limited research has examined the specific mechanisms through which reflection interacts with AI-generated hints in learning environments~\citep{DBLP:conf/iticse/MargulieuxPRBUL24,DBLP:conf/icer/MarwanWP19,DBLP:conf/sigcse/PrasadS24,DBLP:conf/lats/KumarXLMSWLWRSL24,singh2024bridging}. Understanding the underlying mechanism is crucial to ensure reflection intervention is effectively implemented alongside AI-generated hints in real-world classrooms.

\begin{figure}[t!]
    \centering
    \begin{subfigure}{\linewidth}
    {
        \includegraphics[height=0.07\paperheight]{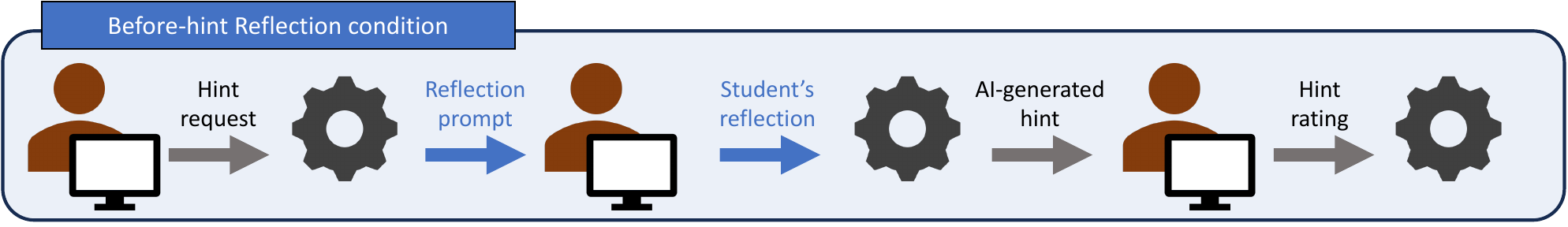}
        \vspace{1.0mm}
        \\
        \includegraphics[height=0.07\paperheight]{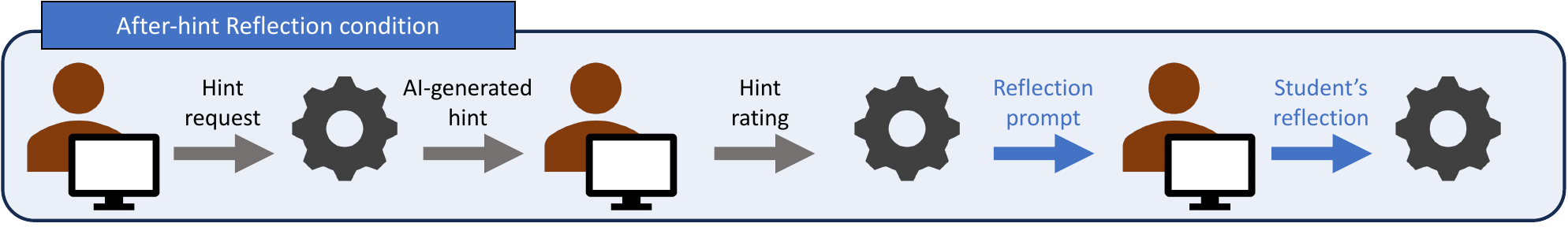}
        \vspace{1.0mm}
        \\
        \includegraphics[height=0.07\paperheight]{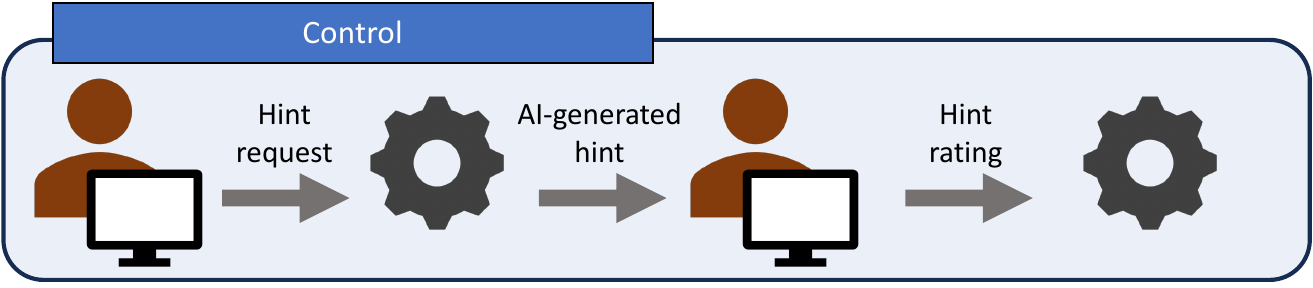}
        \subcaption{Student interactions with the support system in Trial 1.}
        \label{fig:interation_workflow.trial1}
    }
    \end{subfigure}
    \begin{subfigure}{\linewidth}
    {
        \vspace{3.5mm}
        \includegraphics[height=0.07\paperheight]{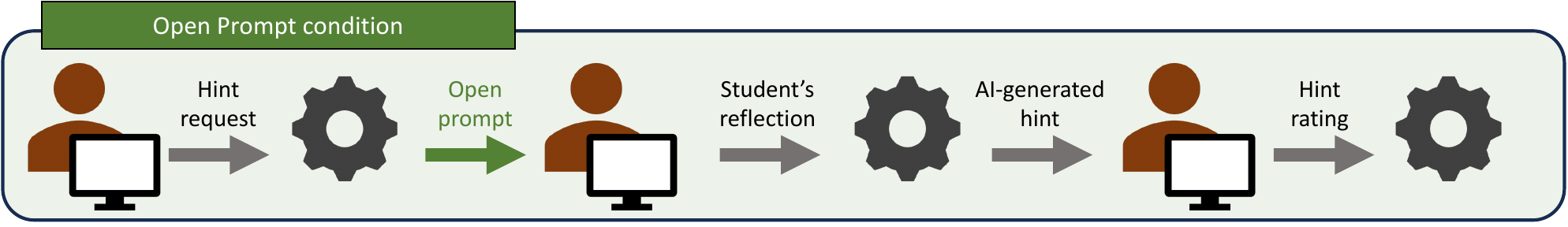}
        \vspace{1.0mm}
        \\
        \includegraphics[height=0.07\paperheight]{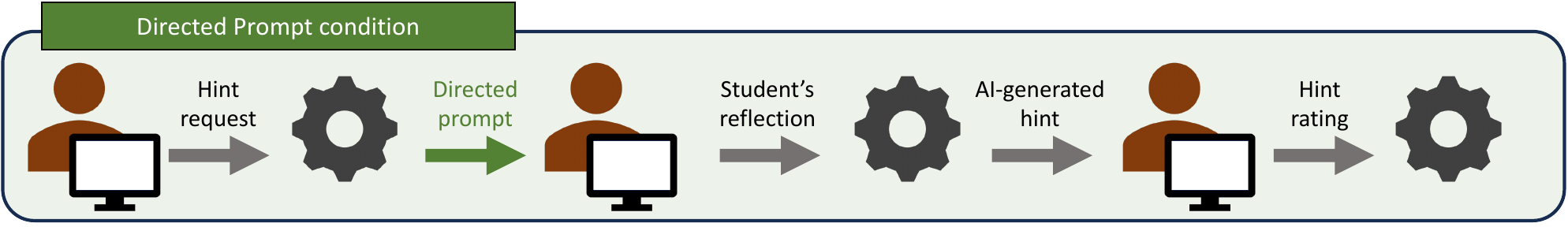}
        \subcaption{Student interactions with the support system in Trial 2.}
        \label{fig:interation_workflow.trial2}
    }
    \end{subfigure}
    \caption{
        Illustration of the two randomized controlled trials involved in our study. Subfigure (a) shows student interactions with the support system in Trial 1.
        Students were randomly assigned to one of three conditions. The Before-hint and After-hint Reflection conditions were prompted for reflection, differing in timing relative to receiving hints. The Control condition only receives hints without reflection prompts. In the Before-hint and After-hint Reflection condition, each time a student requested a hint, one of the three prompts targeting different SRL phases (planning, monitoring, or evaluation) was randomly selected to show to them. See Table~\ref{fig:table_reflection_questions} for all prompts and Table~\ref{fig:pre_post_hint_reflections} for more examples of students' reflections corresponding to different prompts in Trial 1. Subfigure (b) shows student interactions with the support system in Trial 2. Students were randomly assigned to one of two conditions: Open Prompt and Directed Prompt. Depending on the condition, a student was always shown one corresponding reflection prompt whenever they requested a hint. See Table~\ref{fig:table_reflection_questions_study2} for the two reflection prompts and Table~\ref{tab:prompt_type_reflections} for more examples of students' reflections in Trial 2.
    }
    \label{fig:interation_workflow}
\end{figure}

To address this gap, we conducted two field experiments to examine the impact of reflection practices on student metacognitive laziness during interactions with AI-generated hints in an online introductory data science programming course (see Figure~\ref{fig:interation_workflow}). The study was conducted using a system that provided AI-generated hints on demand within students' assignment environment (JupyterLab~\citep{jupyterlab}). The first of these trials focused on understanding the effect of the placement and targeted SRL phases of reflection prompts in the help-seeking process. The second trial focused on understanding how the amount of guidance in reflection prompts affects student behaviors and satisfaction with provided hints. Taken together, these trials address the following three research questions:
\begin{itemize}
\item \textbf{RQ1:} How does the placement of reflection prompts (\emph{before} vs. \emph{after} receiving an AI-generated hint) affect students' hint-seeking experiences, problem-solving performance, and reflective responses?
\item \textbf{RQ2:} How does the SRL phase emphasized in reflection prompts (\emph{planning}, \emph{monitoring}, and \emph{evaluation}) affect students' hint-seeking experiences, problem-solving performance, and reflective responses?
\item \textbf{RQ3:} How does the amount of guidance in reflection prompt (\emph{directed}, providing specific guidance on what to address in a reflection, vs. \emph{open}, where minimal guidance is provided) affect students' hint-seeking experiences, problem-solving performance, and reflective responses?
\end{itemize}

Through these trials, we discovered several important relationships between hint-seeking behaviors, reflective writing, and AI hint generation. Firstly, across trials, we found a consistent inverse relationship between student satisfaction with AI-generated hints and the quality of the students' reflective responses. In addition to that general trend, students who received reflection prompts \emph{before} receiving AI-generated hints produced higher-quality reflections but reported lower satisfaction with the hints compared to those who were prompted \emph{after} hints or not prompted (RQ1). \emph{Planning}-focused reflection prompts generally led to higher-quality reflections and lower satisfaction than \emph{monitoring}- or \emph{evaluation}-focused prompts (RQ2). \emph{Directed} prompts yielded higher-quality reflections and lower hint satisfaction than \emph{open} prompts (RQ3).

The consistent findings of an inverse relationship between satisfaction over AI-generated hints and reflection quality point to a broader pedagogical tension between optimizing AI systems for immediate user satisfaction and the goal of fostering deeper and effortful learning. This tension invites educational stakeholders to reconsider foundational systemic aspects of higher education structures in a time where AI can outperform traditional educational support roles for learners. Specifically, this work contributes to the broader discussions on learning with AI in several ways, as it:

\begin{itemize}
\item Demonstrates the tension between AI systems optimization for user satisfaction and the pedagogical goal for effortful learning, highlighting the need to prioritize pedagogical values over satisfaction in the design and development of educational AI.
\item Provides empirical evidence from an authentic learning environment on how reflective scaffolding during students' interaction with AI-generated hints shapes their hint-seeking experiences, reflection quality, and performance.
\item Underscores the importance of purposefully designing AI-powered educational tools to actively cultivate students’ SRL skills, the key to critical thinking.
\end{itemize}

In summary, this work advances understanding of how reflection and AI-generated hints can be integrated to foster more meaningful learning, and provides a foundation for rethinking how learning systems and AI models might evolve in response to new educational realities.


\section{Related Work}

\subsection{Learning with Reflection}

Reflection is widely recognized as a process of critically reviewing one's learning experiences to strengthen one's comprehension and enrich insights~\citep{Dewey1997-sp,boud1985reflection,boud2013enhancing,ORourke1998-zb}. Studies have demonstrated the metacognitive, cognitive, and emotional benefits of reflection in programming education~\citep{bardach2021power, DBLP:journals/iahe/ChoiJPBJW23}. \citet{DBLP:journals/iahe/ChoiJPBJW23} prompted students to reflect after completing programming assignments to encourage more pedagogically meaningful engagement with assignment hints. The authors found that students who received reflection prompts reported significantly higher perception of learning and performed better on both immediate and delayed post-tests. \citet{DBLP:conf/lats/KumarXLMSWLWRSL24} conducted randomized controlled trials to investigate the effect of post-lesson reflection activities guided by a GPT-3 large language model in an undergraduate computer science course. The authors found that the reflection group not only reported an increase in self-confidence but also outperformed the control group on post-tests after two weeks. \citet{lishinski2021self} found that undergraduate students who completed self-evaluation tasks that targeted reflection and monitoring outperformed their control group in computer science project scores.

Historically, researchers have actively explored how to design reflection prompts to enhance the benefits of reflection. In this section, we focus on three key aspects of reflection prompts: the placement of the prompt (\emph{before} and \emph{after}), its alignment with SRL phases (\emph{planning}, \emph{monitoring}, and \emph{evaluation}), and the amount of guidance it provides (\emph{directed} and \emph{open})~\citep{bardach2021power, DavisLinn2000, mann2009reflection}.

\subsubsection{Placement} 
\label{subsubsection_placement}
Reflection can take place during any stage of learning and activates different thinking skills. Specifically, reflection before action aims to plan for better future consequences. During this prospective reflection, students typically review relevant prior experience or external resources, identify optimal strategies, and simulate their actions~\citep{schon2017reflective, mann2009reflection, taguma2019oecd}. 

On the other hand, reflection after action typically involves self-evaluation. Learners consider the strategies they used and the results they achieved to identify what worked and what could be improved. For example, \citet{loksa2020investigating} showed that one student reflected on adding a visualization and realized that it easier than expected and improved the appearance of their project.

\subsubsection{Self-Regulated Learning Phases}
Reflection is often considered as the final phase of the SRL cycle, where learners review and evaluate their performance after completing a learning task. Zimmerman's cyclical model describes SRL as unfolding across three interrelated phases: forethought (planning), performance (monitoring), and self-reflection (evaluation).~\citep{zimmerman2002becoming}. 
However, reflection is not limited to the final evaluation stage of learning. Reflection can occur throughout all stages of learning (See \ref{subsubsection_placement}). When reflection becomes a part of planning, monitoring, and evaluation phases, it helps students foster metacognitive awareness, adaptive learning strategies, and deeper engagement with learning activities. A growing body of research in programming education shows reflection’s value at every stage.

Students who engage in reflection before starting to code (reflection-before-action, reflection-for-action) tended to develop a deeper understanding of core concepts. For example, when students collaboratively reflected on similar worked examples before their task, they generally learned more conceptual knowledge than those who did not reflect and immediately began problem-solving~\citep{sankaranarayanan2022collaborative}. The study findings also showed that spending more time on reflective planning did not hurt their problem-solving performance.

As a part of the monitoring phase (reflection-in-action), students reflect on their previous practices and test out their hypotheses to decide their next action~\citep{schon2017reflective}. For instance, \citet{Edwards2004-ma} designed an intervention that encouraged students to reflect during software testing activities. The author found that those students produced 45\% fewer bugs per 1000 lines of code. The intervention also increased their confidence when making changes to their code. \citet{Zarestky2022-gl} demonstrated that students who wrote reflection journals as watching lecture videos outperformed their control-group peers.

After task completion, reflective interventions are most often implemented as self-evaluation activities (reflection-on-action) as mentioned in \ref{subsubsection_placement} ~\citep{schon2017reflective}. For example, \citet{lishinski2021self} examined the impact of self-evaluation exercises in an introductory programming course and found that students who participated in these activities earned significantly higher project scores.

\subsubsection{Amount of Guidance}
The reflection prompt, which directs the learner to the reflective writing task, can provide different degrees of guidance. For instance, open prompts encourage students to reflect in ways that are not tied to specific contexts, allowing for greater autonomy and flexibility. These prompts are often open-ended in nature, such as ``As of now, we are thinking of…'', where students are invited to complete the prompt, setting their own direction for the reflection. Open prompts can help to foster personal meaning-making and deeper engagement, especially for learners who are self-motivated or have developed metacognitive skills~\citep{davis2003prompting}. In contrast, directed prompts provide more explicit guidance, suggesting productive directions based on established principles of cognition and instruction, such as ``Looking back on this project, are there specific areas where you wish you had spent more time? If so, which ones?''~\citep{DavisLinn2000, davis2003prompting, rum2017metocognitive}. Directed prompts act as a stronger scaffold of the reflection activity, supporting students who may be less experienced with reflective practice or who benefit from additional examples when organizing their thoughts~\citep{DavisLinn2000,johnson2010applying}.

The effectiveness of open versus directed prompts has been the subject of considerable research and debate. Some studies have found that open prompts can foster more cohesive thought, deeper integration of knowledge, and greater expansion of ideas, particularly among students with higher autonomy~\citep{DavisLinn2000}. Other research suggests that directed prompts are especially effective when combined with structured activities designed to support project completion or specific learning goals~\citep{johnson2010applying}. 

Taken together, these underexplored questions and mixed findings in the literature highlight the importance of considering the placement, associated SRL phases, and the amount of guidance in the reflection prompt when designing interventions to support effective reflection practices of students.

\subsection{AI and AI-generated Hints in Programming Education}
The integration of generative AI into programming education has attracted substantial interest, with numerous studies exploring the potential benefits and challenges of this technology~\citep{DBLP:conf/sigcse/ShenARJS24,DBLP:conf/icer/PhungPCGKMSS22,DBLP:conf/iticse/Prather00BACKKK23,DBLP:conf/sigcse/BeckerDFLPS23}. 
Research on student-AI interaction has found that students are generally receptive to AI support and willing to incorporate it into their studies~\citep{DBLP:conf/aied/MaCK24,valova2024students}. This widespread interest highlights the importance of a timely investigation into the impacts that AI brings to learning environments. 

Existing studies on the impact of AI on learning, however, have shown mixed results. Concerningly, ~\citet{kosmyna2025your} reported that over a four-month period, subjects who used generative AI for essay writing exhibited a substantial decline in brain connectivity and reduced engagement in alpha and beta networks, relative to peers who did not use AI. Regarding AI-generated hints, ~\citet{DBLP:conf/sigcse/WangMP24} reported an encouraging finding that AI-generated hints reduce students' repetitions of errors. On the other hand, research from~\citet{bastani2024generative} and ~\citet{zamfirescu202561a} found no significant learning gains from AI support. \citet{DBLP:journals/corr/abs-2307-00150} noted that when AI-generated hints were no longer available, students who had previously received such hints had lower first-attempt success rates compared to the control group. These findings emphasize the need for further research on mitigating the over-assistance of AI in educational contexts to benefit students' fundamental learning skills in long term. Addressing this need, our work investigates the use of reflection prompts in a learning environment where AI-generated hints are available, examining the interplay between these two forms of support and how reflection prompts can be effectively employed in this setting.

\subsection{Interplay of Reflection and AI-generated Hints}
The intersection between AI-generated hints and reflection in education is a rapidly developing area. However, existing work examined only the correlation between AI usage and self-regulation, without investigating how students' use of AI-generated hints might influence or be influenced by their reflective practices. For instance, Margulieux et al.~\citep{DBLP:conf/iticse/MargulieuxPRBUL24} explored the relationship between AI usage and students' self-regulation, finding no correlation between these elements. This leaves open questions about the dynamic relationship between reflection and AI-generated support, such as how engagement with AI-generated hints could shape students' reflective practices, and vice versa.

To address these gaps, we investigate the influences between reflection and AI-generated hints. Specifically, we examine how the availability of reflection prompts affects students' AI-generated hint uses and how variations in reflection practice (prompt placement, targeted SRL phases, and amount of guidance in the prompt) influence students' overall experiences with AI-generated hints, problem-solving performance, and reflection practices. 


\section{Study Design} \label{sec:study_designn}

This study involved two randomized controlled trials (See Figure~\ref{fig:interation_workflow}), both used a custom support system that can deliver reflection prompts and AI-generated hints. The trials were conducted in two separate offerings of the same online programming course. Each trial examined different factors influencing students' reflection practices, as well as the effects of these practices on learning behaviors.

\begin{figure}[t!]
    \includegraphics[width=\linewidth]{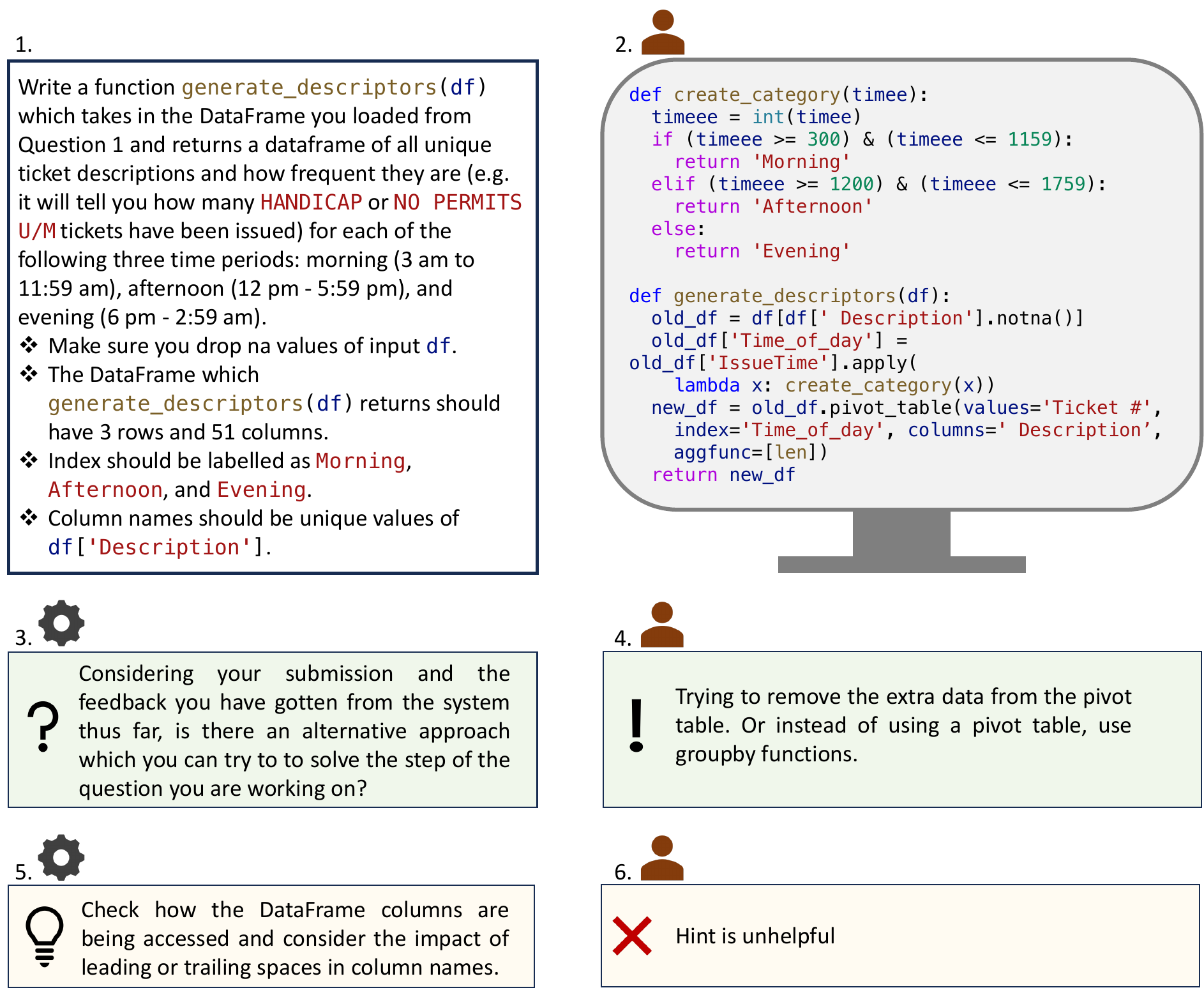}
    \caption{
         Example illustrating a course question and an interaction between a student in the Before-hint Reflection condition and our support system in Trial 1. The figure shows: 1. Problem description, 2. The student's current buggy program, 3. The system's reflection prompt targeting the evaluation SRL phase, 4. The student's reflection, 5. The AI-generated hint, and 6. The student's evaluation of the hint's helpfulness.
    }
    \label{fig:illustration_pre_293}
\end{figure}

Trial 1 investigated the impact of reflection prompts (\emph{presence} vs. \emph{absence}), the placement of prompts (\emph{before} vs. \emph{after} a hint), and the SRL phase targeted by the prompt (\emph{planning} vs. \emph{monitoring} vs. \emph{evaluation})~\citep{zimmerman2002becoming} (See Figure~\ref{fig:illustration_pre_293} for an example). Trial 2 focused on the impact of prompt guidance type (\emph{directed} vs. \emph{open}).

\subsection{Course Context} \label{sec:study_design.study_context}
This study was conducted in two runs of the same introductory data science course using Python, which took place in May 2024 and August 2024. This course is the first mandatory programming course in an online Master's degree program in applied data science at the University of Michigan. Spanning four weeks, the course included weekly assignments, each provided in the form of a JupyterLab notebook containing three to four programming questions, for a total of fourteen questions. Each question required students to complete or modify a given template code to achieve a specific task. Topics covered included regular expressions and dataframe operations using the `\textit{pandas}' library. Students could submit assignments multiple times before the weekly deadlines, and their highest assignment scores were counted toward their final grade. Thus, towards the end, almost all students scored 100\% on most assignments. The students came from diverse backgrounds but were required to possess basic knowledge of Python programming and statistics. There were 74 students enrolled in the course in the May 2024 semester (Trial 1) and 102 students in the August 2024 semester (Trial 2).

\subsection{AI-generated Hints}
\label{subsec:AI-powerd-hints}

We implemented a technique to automatically generate debugging hints using generative AI, with minor modifications between the two runs of the study. The main idea of the technique was adapted from literature~\citep{DBLP:conf/lak/PhungPS0CGSS24,neurips2023gaied_32_zamfirescu-pereira}. It comprises two main steps. First, the system collected additional relevant information about students' bugs. This involves (1) obtaining the buggy output by executing the student's program using the Python interpreter and (2) generating a repaired version of the program by querying a generative AI model. To ensure reliability, the technique verified whether the repaired program correctly solves the task; if it does not, that repaired program was discarded and excluded from the next step. Second, this newly obtained information, together with the student's buggy program, is inserted into a prompt to query the AI model for a personalized hint. Throughout this process, we used GPT-4 as the AI model.

Similar to the original technique in the literature, the first deployment run (Trial 1) did not incorporate students' reflections into the hint-generation prompt. However, we observed that some hints in Trial 1 were misaligned with students' main concerns, as expressed in their reflections, reducing hints' relevance and quality. Consequently, in the second run (Trial 2), we included students' reflections (when available) in the prompt to help the AI model focus on the correct issues and produce more targeted, relevant hints. We also made some updates to the language and format of the prompt in the second run. For example, since OpenAI introduced the \emph{system message} feature to better establish context for the query, we modified our prompts to make use of this capability.

\begin{table}[t]
    \caption{
        Reflection prompts in Trial 1. For the two experimental conditions, the prompts were randomized per hint request instead of per student: A student could receive different prompts in different requests. There was no prompt for the Control condition.
    }
    \label{fig:table_reflection_questions}
    \vspace{-3mm}
    \centering
    \scalebox{0.8}{
        \setlength\tabcolsep{6pt}
        \renewcommand{\arraystretch}{1.3}

\begin{tabular}{l|l|l}
    \hline
    \multicolumn{1}{p{0.1\linewidth}|}{\textbf{Condition}} & 
    \multicolumn{1}{l|}{\textbf{Targeted SRL Phase}} & 
    \multicolumn{1}{p{0.67\linewidth}}{\textbf{Reflection Prompt}} 
    \\
    \hline
    & 
    \multirow{4}{*}{Planning} & 
    \multicolumn{1}{p{0.67\linewidth}}{Considering your submission and the feedback you have gotten from the system thus far, what are the steps you think must be followed in order to answer this question, and which step is the one you are currently stuck on?}
    \\
    \cline{2-3}
    \multicolumn{1}{p{0.15\linewidth}|}{\phantom{} \newline Before-hint Reflection} &
    \multirow{3.5}{*}{Monitoring} &
    \multicolumn{1}{p{0.67\linewidth}}{Considering your submission and the feedback you have gotten from the system thus far, which topics in the course do you think are most relevant to the current problem you are facing?}
    \\
    \cline{2-3}
    &
    \multirow{3.5}{*}{Evaluation} &
    \multicolumn{1}{p{0.67\linewidth}}{Considering your submission and the feedback you have gotten from the system thus far, is there an alternative approach which you can try to solve the step of the question you are working on?}
    \\
    \hline
    & 
    \multirow{2.5}{*}{Planning} & 
    \multicolumn{1}{p{0.67\linewidth}}{Considering the hint you just received and your solution thus far, what steps will you take next to move forward on the question?}
    \\
    \cline{2-3}
    \multicolumn{1}{p{0.15\linewidth}|}{\phantom{} \newline After-hint Reflection} &
    \multirow{3.25}{*}{Monitoring} &
    \multicolumn{1}{p{0.67\linewidth}}{Considering the hint you just received and your solution thus far, are there other topics from the course material you should be incorporating into your solution?}
    \\
    \cline{2-3}
    &
    \multirow{4}{*}{Evaluation} &
    \multicolumn{1}{p{0.67\linewidth}}{Considering the hint you just received and your solution thus far, was your general approach a good one, or should you change to an alternative approach to solve the step of the question you are working on?}
    \\
    \cline{2-3}
    \hline
    Control & N/A & N/A
    \\
    \hline
\end{tabular}

    }
\end{table}

\begin{table}[t]
    \caption{
        Reflection prompts in Trial 2. Different from Trial 1, in this trial, the prompts were randomized per student: a student always received the same prompt across all hint requests.
    }
    \label{fig:table_reflection_questions_study2}
    \vspace{-3mm}
    \centering
    \scalebox{0.8}{
        \setlength\tabcolsep{9pt}
        \renewcommand{\arraystretch}{1.3}

\begin{tabular}{l|l}
    \hline
    \multicolumn{1}{l|}{\textbf{Condition}} & 
    \multicolumn{1}{p{0.7\linewidth}}{\textbf{Reflection Prompt}} 
    \\
    \hline
    \multirow{1.7}{*}{Open} & \multicolumn{1}{p{0.96\linewidth}}{Considering the program you wrote and the feedback you have received from the system so far, what do you think is a possible bug in the program?}
    \\
    \hline
    \multirow{3.25}{*}{Directed} & \multicolumn{1}{p{0.96\linewidth}}{Considering the program you wrote and the feedback you have received from the system so far, what do you think is a possible bug in the program? How does the bug affect the program? What do you think is a way to fix the bug?}
    \\
    \hline
\end{tabular}
    }
\end{table}

\subsection{Reflection Prompts}
\label{subsec:reflection_practice}

In Trial 1, students were randomly assigned to one of three conditions: Before-hint Reflection, After-hint Reflection, or Control (See Figure~\ref{fig:interation_workflow.trial1}). Each interaction began when a student clicked the hint-request button for an assignment question. In the Before-hint Reflection condition, students reflected on their progress before receiving a hint (reflection-for-action). In the After-hint Reflection condition, students reflected after receiving and rating a hint (reflection-on-action). The Control group received hints without any reflection prompt.

Three reflection prompts were developed for each experimental condition, with each prompt targeting a SRL phase: planning, monitoring, or evaluation~\citep{zimmerman2002becoming}. Planning prompts asked students to outline their approach; monitoring prompts encouraged them to connect the task to course material; and evaluation prompts promoted self-assessment and consideration of alternatives (See Table~\ref{fig:table_reflection_questions}). These prompts were randomized by hint request, not by student, to examine how prompt type impacted students’ needs during problem-solving. For example, a student could receive a planning prompt for one hint request and an evaluation prompt for another.

In Trial 2, students were randomly assigned to one of the two conditions: Open Prompt or Directed Prompt (See Figure~\ref{fig:interation_workflow.trial2}). Depending on their condition, a student was always shown one of the two prompt types (see Table~\ref{fig:table_reflection_questions_study2}). The open prompt asked students to reflect on their progress in general, while the directed prompt guided them to reflect on more specific aspects, such as the impact of the bug on their program and possible solutions.

\subsection{Instruments and Data Collection} \label{subsec:study_design.instruments}

Our support system consisted of a front-end JupyterLab extension to interact with the students and a back-end server application to execute the hint-generation technique detailed in Section~\ref{subsec:AI-powerd-hints}.

In Trial 1, students needed to click on an Activation button on their JupyterLab and agree to a consent notice to activate our support system. Upon activation, the extension added a "Hint" button below each assignment question. When clicked, it queried the back-end, which generated and returned a hint. Then the system displayed the hint at the top of the screen, providing an option for the student to rate it as either Helpful or Unhelpful. Depending on their condition, students were also prompted to write a reflection, as detailed in Section~\ref{subsec:reflection_practice} and illustrated in Figure~\ref{fig:interation_workflow.trial1}. This reflection practice was optional, and students were allowed to leave the field blank. To prevent excessive use of hints, students were limited to three hint requests per question.

In Trial 2, we made a subtle modification on the front-end: The hint button was shown directly from the beginning, and students could agree to the consent notice and activate hintbot when they request the first hint. This update is aimed at making the system more visible and reducing friction. We still maintain a quota for the number of hints per question, but increase the limit to five hints to accommodate students' needs. In this trial, all prompts were served before the students received a hint, regardless of the condition. The rest of the interaction workflow remained the same as in Trial 1.

Interacting with the support system was optional, and students were informed that the hints were generated by AI and might not be accurate. We also notified that we collected data on their interactions with the support system, including the timestamps of hint requests, reflection prompts shown, students' reflections, hints provided, their ratings of hints, along with their assignment submissions. Tables~\ref{fig:participation} and \ref{fig:participation_study2} outline the number of students and their hint requests in our two trials, respectively.

\begin{table}[t!]
    \caption{Overview of student participation in Trial 1. Subfigure (a) shows statistics regarding the students: The first row shows the number of students who activated the support system. The second row shows the number of students who have completed at least one interaction with the support system (i.e., receiving a hint for the Control condition or a hint-prompt pair for the experimental conditions). Subfigure (b) shows statistics regarding the hint requests for different conditions and reflection prompts. For space, the SRL phases of planning, monitoring, and evaluation are abbreviated as Plan., Moni., and Eval., respectively. We note that for the After-hint Reflection condition, some hints (8) are not associated with any prompt ($\varnothing$) because in those interactions, students did not rate the hints and thus, were not presented with any prompt. Regarding the second row, an interaction is completed if the student has received both a hint and, for the experimental conditions, a reflection prompt.}  
    \label{fig:participation}
    \vspace{-1mm}
    \begin{subtable}{\linewidth}
    {
        \subcaption{Statistics regarding the students}
        \label{fig:participation.students}
        \scalebox{0.8}{
            \setlength\tabcolsep{3.9pt}
            \renewcommand{\arraystretch}{1.3}
\begin{tabular}{lcccc}
    \hline
    \multirow{2}{*}{\textbf{Statistics}} & \multicolumn{1}{p{0.175\linewidth}}{\textbf{Before-hint Reflection}} & \multicolumn{1}{p{0.165\linewidth}}{\textbf{After-hint Reflection}} & \multirow{2}{*}{\textbf{Control}} & \multicolumn{1}{p{0.1\linewidth}}{\textbf{Grand Total}} \\
    \hline
    Students activated the support system & 12 & 13 & 9 & 34 \\
    Students completed at least one interaction & 12 & 13 & 8 & 33 \\
    \hline
\end{tabular}

        }

    }
    \end{subtable}
    \begin{subtable}{\linewidth}
    {
        \vspace{4mm}
        \subcaption{Statistics regarding the hint requests}
        \label{fig:participation.help_requests}
        \scalebox{0.8}{
            \setlength\tabcolsep{3.5pt}
            \renewcommand{\arraystretch}{1.3}
\begin{tabular}{lccccccccccc}
    \hline
    \multicolumn{1}{l|}{\multirow{2}{*}{\textbf{Statistics}}} & 
    \multicolumn{4}{c|}{\textbf{Before-hint Reflection}} & 
    \multicolumn{5}{c|}{\textbf{After-hint Reflection}} & 
    \multicolumn{1}{c|}{\multirow{1}{*}{\textbf{Control}}} & 
    \multicolumn{1}{c}{\multirow{1}{*}{\textbf{Grand}}} 
    \\
    \multicolumn{1}{l|}{}  &
    \multicolumn{1}{l}{\textbf{Total}} &
    \multicolumn{1}{l}{Plan.} &
    \multicolumn{1}{l}{Moni.} &
    \multicolumn{1}{l|}{Eval.} &
    \multicolumn{1}{l}{\textbf{Total}} &
    \multicolumn{1}{l}{Plan.} &
    \multicolumn{1}{l}{Moni.} &
    \multicolumn{1}{l}{Eval.} &
    \multicolumn{1}{l|}{$\varnothing$} &
    \multicolumn{1}{c|}{\textbf{Total}} &
    \multicolumn{1}{c}{\multirow{1}{*}{\textbf{Total}}}
    \\
    \hline
    \multicolumn{1}{l|}{Hints provided} &
    \multicolumn{1}{c}{52} &
    \multicolumn{1}{c}{16} &
    \multicolumn{1}{c}{17} &
    \multicolumn{1}{c|}{19} &
    \multicolumn{1}{c}{56} &
    \multicolumn{1}{c}{17} &
    \multicolumn{1}{c}{11} &
    \multicolumn{1}{c}{20} &
    \multicolumn{1}{c|}{8} &
    \multicolumn{1}{c|}{50} &
    \multicolumn{1}{c}{158}
    \\
    \multicolumn{1}{l|}{Interactions} &
    \multicolumn{1}{c}{52} &
    \multicolumn{1}{c}{16} &
    \multicolumn{1}{c}{17} &
    \multicolumn{1}{c|}{19} &
    \multicolumn{1}{c}{48} &
    \multicolumn{1}{c}{17} &
    \multicolumn{1}{c}{11} &
    \multicolumn{1}{c}{20} &
    \multicolumn{1}{c|}{0} &
    \multicolumn{1}{c|}{50} &
    \multicolumn{1}{c}{150}
    \\
    \multicolumn{1}{l|}{Hints rated} &
    \multicolumn{1}{c}{50} &
    \multicolumn{1}{c}{16} &
    \multicolumn{1}{c}{16} &
    \multicolumn{1}{c|}{18} &
    \multicolumn{1}{c}{48} &
    \multicolumn{1}{c}{17} &
    \multicolumn{1}{c}{11} &
    \multicolumn{1}{c}{20} &
    \multicolumn{1}{c|}{0} &
    \multicolumn{1}{c|}{41} &
    \multicolumn{1}{c}{139}
    \\
    \hline
\end{tabular}

        }

    }
    \end{subtable}
    \vspace{2mm}
\end{table}
\begin{table}[t!]
    \caption{Overview of student participation in Trial 2. Similar to Table~\ref{fig:participation}, this table outlines statistics regarding students and hint requests in Trial 2.}  
    \label{fig:participation_study2}
    \vspace{-1mm}
    \scalebox{0.8}{
        \setlength\tabcolsep{16pt}
        \renewcommand{\arraystretch}{1.3}
\begin{tabular}{lccc}
    \hline
    \multirow{2}{*}{\textbf{Statistics}} & \multicolumn{1}{p{0.12\linewidth}}{\textbf{Open Prompt}} & \multicolumn{1}{p{0.12\linewidth}}{\textbf{Directed Prompt}} & \multicolumn{1}{p{0.09\linewidth}}{\textbf{Grand Total}} \\
    \hline
    Students activated the support system & 53 & 45 & 98 \\
    Students completed at least one interaction & 37 & 27 & 64 \\
    \hline
    Hints provided & 236 & 175 & 411 \\
    Interaction completed & 236 & 175 & 411 \\
    Hints rated & 213 & 168 & 381 \\    
    \hline
\end{tabular}
    }
\end{table}

\section{Analysis Setup} \label{sec:analysis_setup}

This section outlines our data analysis procedures for addressing Research Questions 1–3 through two randomized controlled trials. Findings from the Trial 1 were analyzed to answer RQs 1–2, and those from Trial 2 were used to address RQ3.

\subsection{RQ1: Impacts of Reflection Prompt and Its Placement} \label{sec:analysis_setup.rq1}

We subdivide RQ1 into four sub-RQs as follows: (a) and (b) examine how reflection practices affect students' hint-seeking experience and problem-solving performance, respectively. (c) and (d) examine how prompt placements (before-hint versus after-hint) affect students' participation rates of reflection practices and quality of reflection responses, respectively.

For RQ1a, we measured the hint-seeking experience based on (1) the total number of hint requests made by students throughout the course and (2) their satisfaction with the hints, as indicated by their hint ratings.

For RQ1b, we measured problem-solving performance using two metrics: First, we computed the number of assignment submissions throughout the course. As noted in Section~\ref{sec:study_design.study_context}, by the end of the course, almost all students succeeded in solving all questions. Thus, we used the number of submissions as an indicator of student performance: the fewer the submissions required to achieve their highest scores, the better the performance. Second, we measured performance using the immediate success rate: whether the question was solved in the submission immediately following the hint request.

For RQ1c, we calculated the participation rates for each experimental condition using two metrics: (1) the number of students who wrote at least one non-empty reflection and (2) the number of non-empty reflections.

For RQ1d, we conducted a thematic analysis of students' written responses to before- and after-hint reflection prompts \citep{braun2023thematic}. Thematic analysis is a robust qualitative method for identifying and interpreting themes in textual data, and is well-suited for both inductive and deductive approaches \citep{braun2023thematic, maguire2017doing}. In this study, we adopted a deductive approach focused on describing objective reality \citep{naeem2023pragmatic}, theoretically drawing on cyclical SRL model~\citep{zimmerman2002becoming} and the 5Rs framework for hierarchical levels of reflection. The 5Rs include Reporting (level 1), Responding, Relating, Reasoning, and Reconstructing (level 5), and served as theoretical grounding to guide the coding and interpretation of student reflections \citep{Bain1999-ji,Bain2002-qc,ryan2014reflection}. According to the 5Rs framework, higher levels of reflection move beyond basic description to the integration of theory and experience, and enables learners to explain, interrogate, and ultimately transform their practice \citep{ryan2014reflection}. This progression is closely aligned with the development of autonomous learning practices and critical thinking.

The qualitative analysis proceeded in an iterative manner to ensure rigor and reliability. In the first iteration, two authors with strong familiarity with the course independently performed open coding on 10\% of the student reflection data, guided by the SRL model~\citep{zimmerman2002becoming} and the reflection level framework~\citep{Bain1999-ji,Bain2002-qc,ryan2014reflection}. After initial coding, these authors met to discuss their results, collaboratively developed a draft codebook, and identified broader themes. In the second iteration, another author joined the coding team as the third coder. All three coders then independently analyzed the remaining data using the draft codebook, while staying open to new codes as they emerged. For the final step, the three authors worked together to refine the draft codebook (see Table~\ref{fig:table_codes}). They systematically reviewed the coded data and resolved any discrepancies. This collaborative process ensured both consensus and consistency in the identification of themes across all student reflections. The thematic analysis approach provided a nuanced understanding of how prompt placement influences the depth and quality of student reflection.

\begin{table}[t!]
    \caption{
        Qualitative codes for thematic analysis of students' reflections in Trial 1. The creation of these codes is outlined in Section~\ref{sec:analysis_setup.rq1}.
    }
    \label{fig:table_codes}
    \vspace{-1mm}
    \centering
    \scalebox{0.84}{
        \setlength\tabcolsep{8pt}
        \renewcommand{\arraystretch}{1.2}

\begin{tabular}{l|l}
    \hline
    \multicolumn{1}{l|}{\textbf{Theme}} & 
    \multicolumn{1}{p{0.8\linewidth}}{\textbf{Code and Description}} 
    \\
    \hline
    & 
    \multicolumn{1}{p{0.8\linewidth}}{Planning (students plan how to approach the problem or set goals for learning)}
    \\
    \cline{2-2}
    \multirow{1}{*}{SRL Phase} &
    \multicolumn{1}{p{0.8\linewidth}}{Monitoring (students monitor their progress while working on the problem)}
    \\
    \cline{2-2}
    &
    \multicolumn{1}{p{0.8\linewidth}}{Evaluation (students evaluate their performance)}
    \\
    \hline
    \multirow{4}{*}{Critical Engagement} & 
    \multicolumn{1}{p{0.8\linewidth}}{What (Describing key issues/events)}
    \\
    \cline{2-2}
    &
    \multicolumn{1}{p{0.8\linewidth}}{Why (Explaining why key issues/events are important)}
    \\
    \cline{2-2}
    &
    \multicolumn{1}{p{0.8\linewidth}}{How (Discussing relevant knowledge they need to resolve issues)}
    \\
    \hline
    & 
    \multicolumn{1}{p{0.8\linewidth}}{Regular expression (including format/pattern matching, quantifier, lookaheads, list comprehension, \textit{VERBOSE})}
    \\
    \cline{2-2}
    \multirow{4}{*}{Course Contents} &
    \multicolumn{1}{p{0.8\linewidth}}{\textit{pandas} (including pivot table, \textit{groupby}, sort, \textit{dropna}, \textit{merge}, \textit{concat}, 
    \textit{filter}, array, dataframe, column name change, find the right column)}
    \\
    \cline{2-2}
    &
    \multicolumn{1}{p{0.8\linewidth}}{\textit{NumPy} (\textit{ndarray})}
    \\
    \cline{2-2}
    &
    \multicolumn{1}{p{0.8\linewidth}}{\textit{Python} (\textit{for} loop)}
    \\
    \cline{2-2}
    &
    \multicolumn{1}{p{0.8\linewidth}}{Math (percentage)}
    \\
    \hline
    \multirow{3}{*}{Hint Assessment} & 
    \multicolumn{1}{p{0.8\linewidth}}{Helpfulness (including the hint is helpful or unhelpful)} 
    \\
    \cline{2-2}
    &
    \multicolumn{1}{p{0.8\linewidth}}{Relevancy (including the hint is relevant or irrelevant)}
    \\
    \cline{2-2}
    &
    \multicolumn{1}{p{0.8\linewidth}}{Same as the previous hint}
    \\
    \hline
\end{tabular}

    }
    \vspace{2mm}    
\end{table}

\subsection{RQ2: Impacts of SRL Phase Associated with Reflection Prompts} \label{sec:analysis_setup.rq2}

We divided RQ2 into four sub-questions to examine the effects of prompts associated with different SRL phases, using the measures established for RQ1, including: (a) hint-seeking experience, (b) problem-solving performance, (c) participation rates of reflection practices, and (d) quality of reflection responses.

For this analysis, we omit the After-hint Reflection condition due to a low participation rate of the students in this condition, described more fully in Section~\ref{sec:results.reflection_impacts}. Due to the small data size, we did not perform statistical tests for RQ2.

Since in Trial 1, prompts were randomly presented for each hint request (see Section~\ref{subsec:reflection_practice}), rather than assigned per student throughout the course, the impacts of prompts could only be measured at the request level, not the student level. Therefore, for the hint-seeking experience in RQ2a, we only measured the hint satisfaction rate, but not the number of hints delivered per student, as in RQ1a. Similarly, for RQ2b about problem-solving performance, we examined the immediate success rates but not the number of submissions per student. For the participation rate in RQ2c, we calculated the number of non-empty reflections. Regarding reflection quality for RQ2d, we analyzed the reflections using the expert-coded annotations described in Section~\ref{sec:analysis_setup.rq1}.

\subsection{RQ3: Impacts of Prompt Types in Reflection Practices} \label{sec:analysis_setup.rq3}

RQ3 was addressed through Trial 2. Similar to above, we also subdivided RQ3 into sub-RQs, examining how different types of prompt guidance affect students' (a) hint-seeking experience, (b) problem-solving performance, (c) participation rate of reflection practices, and (d) the quality of reflection responses.

For RQ3a, RQ3b, and RQ3c, we employed the same analytical methods as those applied to RQ1.
For RQ3d, we conducted a qualitative coding of student reflection responses using a refined version of the codebook developed during Trial 1. In the initial review of a randomly selected set of 10 responses, the original two coders found it necessary to adapt the Reflection involving SRL phase and Reflection involving Critical Engagement themes, as well as the codes of these themes, to better fit the specific research questions and programming tasks in this study. Because only before-hint reflection prompts were used in RQ3 of Trial 2, the Hint Evaluation theme and its codes, which were originally developed for after-hint reflections in Trial 1, were omitted (see Table~\ref{tab:reflection_quality_codes} for the refined codebook).

\begin{table}[t]
    \caption{
        Qualitative codes for the theme "Reflection Quality" used in the analysis of student reflections in Trial 2. Each code is listed with its description. We note that only reflections containing original student text were analyzed. Responses made up solely of copied and pasted error messages were not analyzed. Codes were assigned based on explicit statements in the student reflections, and interpretations were limited to what students actually wrote, without inferring additional meaning.
    }
    \label{tab:reflection_quality_codes}
    \vspace{-1mm}
    \centering
    \scalebox{0.84}{
        \setlength\tabcolsep{9pt}
        \renewcommand{\arraystretch}{1.2}
        \begin{tabular}{l|l|p{0.665\linewidth}}
    \hline
    \textbf{Theme} & \textbf{Code} & \textbf{Description} \\
    \hline
    \multirow{12.75}{*}{Critical Engagement}
      & \multirow{4}{*}{What} & Describing the current error: Student identifies or summarizes the error they are facing (e.g., what is not working, what is incorrect). This is an objective account of an issue encountered at the time of reflection. \\
    \cline{2-3}
      & \multirow{4}{*}{Why} & Explaining why the error occurs: Student offers reasons or opinions about the cause of the error or why it made problem-solving challenging. This reflects the student's own understanding, regardless of whether it is correct or not. \\
    \cline{2-3}
      & \multirow{4}{*}{How} & Suggesting how to resolve the error or its cause: Student mentions knowledge, skills, or alternative approaches that could potentially address the error or its root cause. These are strategies the student has not yet attempted at the time of reflection. \\
    \hline
\end{tabular}
    }
    \vspace{2mm}    
\end{table}

After revising the codebook, the same three coders from Trial 1 independently coded a new randomly selected set of 10 responses. Following this round, they discussed and clarified their interpretations of the themes and codes to ensure consistency. Each coder then independently coded 400 responses, after which the coders met to compare their results and reconcile any disagreements. Krippendorff’s Alpha~\citep{krippendorff2011computing} was used to assess inter-rater reliability among the three coders for the qualitative coding of RQ3d. For the three Critical Engagement codes (what, why, and how), the reliability values before resolving disagreements were 0.791, 0.649, and 0.670, respectively. After the resolution of discrepancies, these values increased substantially to 0.977, 0.916, and 0.950, indicating excellent agreement among coders.


\section{Results} \label{sec:results}

This section presents our results following the analysis setup in Section~\ref{sec:analysis_setup}.

\subsection{RQ1: Impacts of Reflection Prompt and Its Placement} \label{sec:results.reflection_impacts}


\begin{figure}[]
    \centering
    \begin{subfigure}{\linewidth}
    {
        \centering
        \scalebox{0.94}{
            \includegraphics[width=0.95\linewidth]{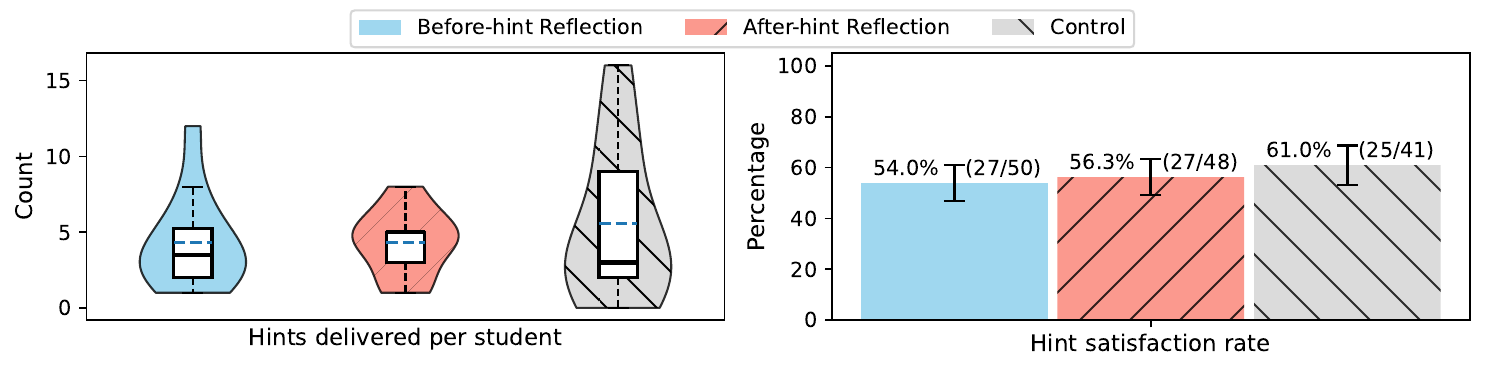}
        }
        \vspace{-2.5mm}
        \subcaption{Results for RQ1a: Impacts of reflection practices on students' hint-seeking experience. Hint-seeking experience is measured by the number of hints requested and delivered to the students (left) and their satisfaction rate with the delivered hints (right). In the left plot, the black horizontal lines show the median and the blue dashed lines show the mean.}
        \label{fig:result_reflection_impacts.help_seeking}
    }
    \end{subfigure}
    \begin{subfigure}{\linewidth}
    {
        \vspace{1.5mm}
        \centering
        \scalebox{0.94}{
            \includegraphics[width=0.95\linewidth]{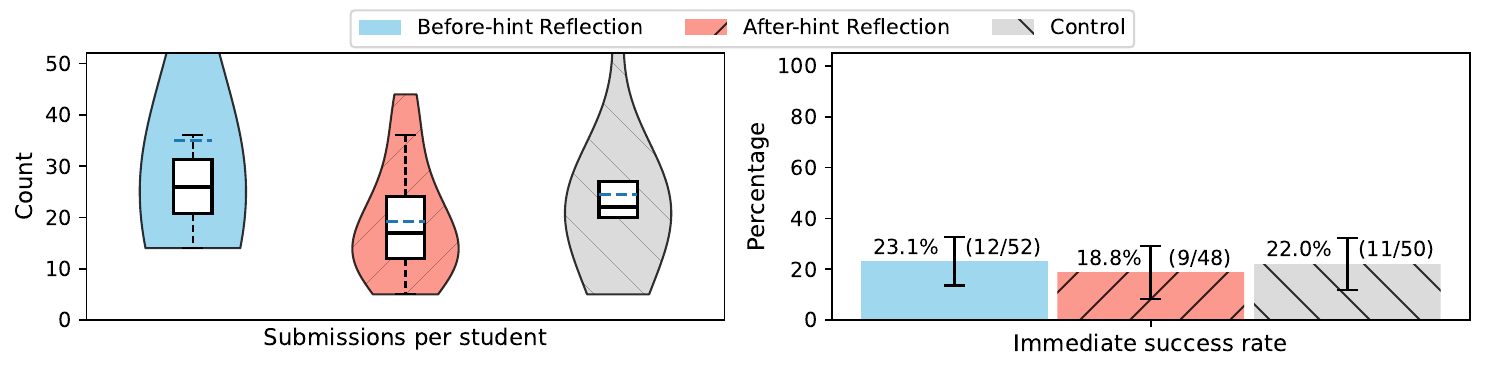}
        }
        \vspace{-2.5mm}
        \subcaption{Results for RQ1b: Impacts of reflection practices on students' solving performance. Solving performance is measured by the total number of submissions made by the students throughout the course (left) and the success rate of solving the questions with the submissions immediately following hint requests (right). In the left plot, the black horizontal lines show median and the blue dashed lines show mean. There is a student in the Before-hint Reflection condition with 146 submissions, and another student in the Control condition with 69 submissions that are not shown due to cutting off of the y-axis for visibility.
        }
        \label{fig:result_reflection_impacts.solving_performance}
    }
    \end{subfigure}
    \begin{subfigure}{\linewidth}
    {
        \centering
        \scalebox{0.94}{
            \includegraphics[width=0.95\linewidth]{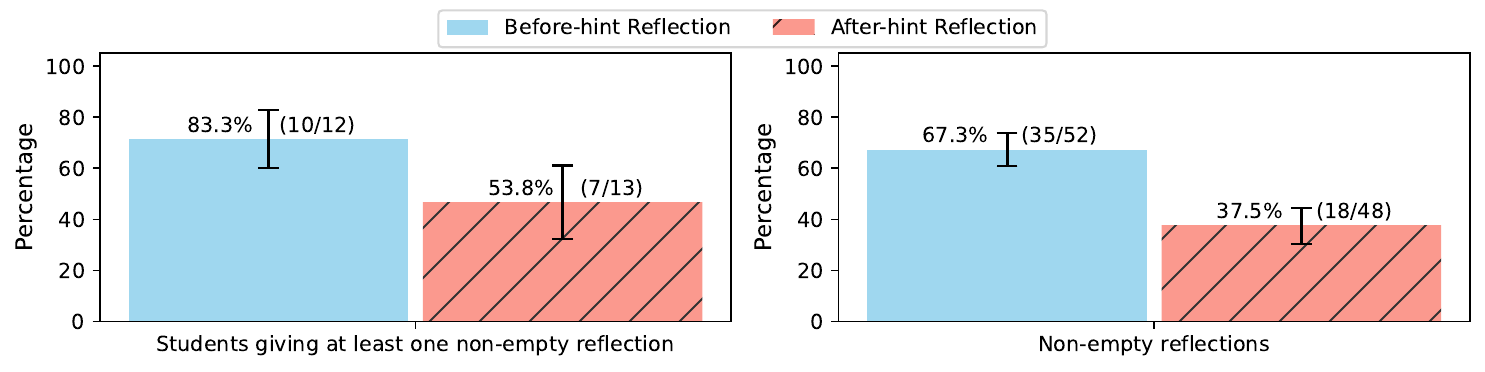}
        }
        \vspace{-2.5mm}
        \subcaption{
            Results for RQ1c: Impacts of prompt placement on students' participation rate in reflection practices. The figure shows the percentages of students who wrote at least one non-empty reflection among all students (left) and the percentages of reflections that are non-empty among all reflections (right).
        }
        \label{fig:result_impact_reflection.participation_rate}
    }
    \end{subfigure}
    \begin{subfigure}{\linewidth}
    {
        \vspace{1.5mm}
        \centering
        \scalebox{0.94}{
            \includegraphics[width=0.95\linewidth]
            {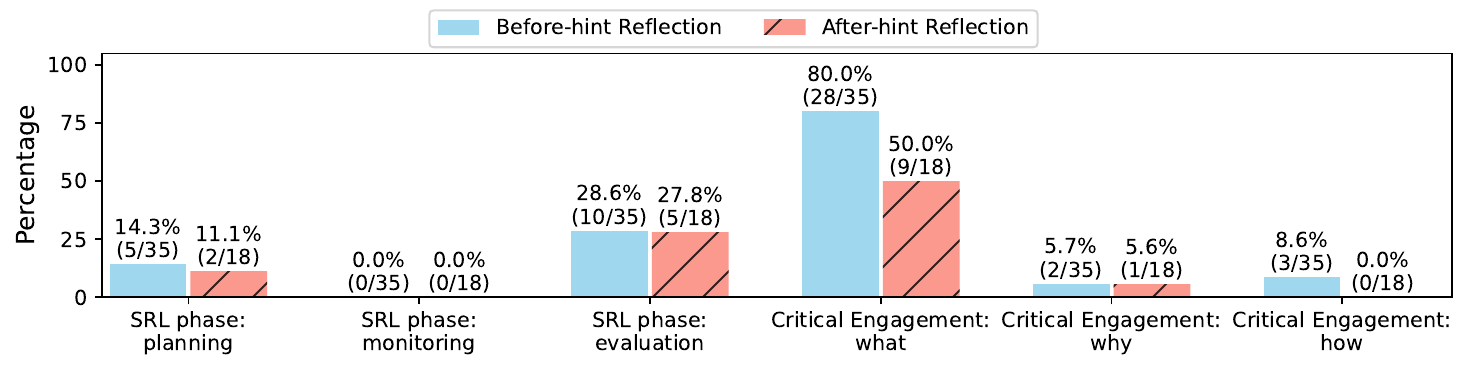}
        }
        \vspace{-2.5mm}
        \subcaption{
            Results for RQ1d: Impacts of prompt placement on students' reflection quality. The figure shows the percentage of non-empty reflections that incorporate different SRL phases and critical engagement, as outlined in Section~\ref{sec:analysis_setup.rq2} and detailed in Table~\ref{fig:table_codes}.
        }
        \label{fig:result_impact_reflection.reflection_qualities}
    }
    \end{subfigure}
    
    \vspace{-2mm}
    \caption{
        Results for RQ1: Impacts of reflection practices and placement.
    }
    \label{fig:result_reflection_impacts}
\end{figure}

\subsubsection{RQ1a: Impact of Reflection Practices on Hint-seeking Experience}
To understand the impact on the hint-seeking experience, we measure the total number of hint requests made by students throughout the course and their satisfaction with the AI-generated hints (see Section~\ref{sec:analysis_setup.rq1}). We calculate the mean number of hint requests per student in each condition. Results indicate a weak relationship. The students in the Control condition requested more hints on average compared to those in the two experimental conditions (see Figure~\ref{fig:result_reflection_impacts.help_seeking} left). However, a two-tailed t-test~\citep{student1908probable} indicates that the difference was not statistically significant ($t=0.92$, $p=0.366$, $df=32$).
An ANOVA test~\citep{girden1992anova} to compare the three conditions also results in no significant difference ($F=0.41$, $p=0.668$, $df1=2$, $df2=31$). It is worth noting that the interquartile ranges of the number of requested hints for the experimental conditions are narrower than that of the Control condition.

Next, we examine hint satisfaction rates across conditions by calculating the percentages of hints rated as Helpful rather than Unhelpful. As shown in Figure~\ref{fig:result_reflection_impacts.help_seeking} right, the Before-hint Reflection condition has the lowest mean satisfaction rating at $54.0\%$, while the Control condition has the highest at $61.0\%$. However, a $\chi^2$ test suggests no statistically significant differences between these three conditions ($\chi^2=0.46$, $p=0.796$, $df=2$).

\subsubsection{RQ1b: Impact of Reflection Practices on Problem-solving Performance}
Problem-solving performance is measured using two metrics: the total number of submissions made and the immediate success rate in solving the questions (see Section~\ref{sec:analysis_setup.rq1}). As shown in Figure~\ref{fig:result_reflection_impacts.solving_performance} left, the Before-hint Reflection condition exhibits the highest number of submissions per student ($35.1$), followed by the Control condition ($24.6$) and the After-hint Reflection condition ($19.2$). An ANOVA test indicates no statistically significant differences across three conditions ($F=1.36$, $p=0.271$, $df1=2$, $df2=31$).

Regarding immediate success, the Before-hint Reflection condition achieves the highest rate ($23.1\%$), closely followed by the Control condition ($22.0\%$), with both being higher than the After-hint Reflection condition ($18.8\%$). A $\chi^2$ test indicates no statistically significant differences across conditions ($\chi^2=0.30$, $p=0.861$, $df=2$).


\subsubsection{RQ1c: Impact of Prompt Placement on Participation Rates}
To understand the impact of placement of reflection prompts with regards to hints, we compare participation rates in reflection between the Before-hint and After-hint Reflection conditions by examining the percentage of students who wrote at least one non-empty reflection and the percentage of non-empty reflections overall. As demonstrated in Figure~\ref{fig:result_impact_reflection.participation_rate}, the Before-hint Reflection condition exhibits higher rates across both metrics. $\chi^2$ tests do not show a significant difference in terms of percentage of students ($\chi^2=2.49$, $p=0.114$, $df=1$) but indicate significant differences in the percentage of non-empty reflections between the conditions ($\chi^2=8.90$, $p=0.003$, $df=1$).

\subsubsection{RQ1d: Impact of Prompt Placement on Reflection Quality}
To understand the impact of reflection prompt placement on reflection practices, we conducted thematic analysis on students' reflection responses to the given prompts (see Section~\ref{sec:analysis_setup.rq1}), found two emerging themes: ``SRL Phases'' and ``Critical Engagement''. A theme of SRL Phase represents with which specific SRL phases students engaged while responding to each reflection prompt designed to activate one of ``planning'', ``monitoring'', and ``evaluation'' phases. Critical Engagement is a theme of critical reflective engagement, which are ``What (Describing the current key issues/events)'', ``Why (Explaining why the issues are important)'', and ``How (Analyzing relevant knowledge they need to resolve the issues).'' 

Thematic analysis (see Section~\ref{sec:analysis_setup.rq1}) revealed clear differences in the content and focus of student reflections between the two experimental conditions (See Table~\ref{fig:pre_post_hint_reflections} for examples of reflections and their qualitative code results). Reflections written before receiving hints (before-hint) tended to be more detailed and substantive, frequently referencing specific course concepts, SRL phases, and practical strategies. For instance, one student described planning how to address a data manipulation challenge: "\textit{Trying to remove the extra data from the pivot table. Or [I can] instead of using a pivot table, use groupby functions}.". Others reflected on the challenges of getting started, with comments like "\textit{Right now I need to formulate the problem better I think.}" and "\textit{where to start}" (See Table~\ref{fig:pre_post_hint_reflections.pre}).

\begin{table}[t!]
     \caption{
        Examples of reflections written by students in the Before-hint and After-hint Reflection conditions (in Trial 1) with annotation of SRL-Phases and Critical Engagement (Crit. Eng.). The expert annotation process is described in Section~\ref{sec:analysis_setup.rq1} and the complete list of codes is shown in Table~\ref{fig:table_codes}. The assignment (A) and question (Q) number followed every response. Coding labels are abbreviated. For SRL phases, P stands for planning, M for monitoring, and E for evaluation. For Critical Engagement, D stands for description of key issues or events (what), E for explanation of why the key issues or events are important (why), and A for analysis of relevant knowledge they need (how).
    }
    \label{fig:pre_post_hint_reflections}
    \vspace{-2.5mm}
    \centering
    \begin{subtable}{\linewidth}
        \centering
        \subcaption{Before-hint Reflection condition}
        \label{fig:pre_post_hint_reflections.pre}
        \vspace{-1.5mm}
        \scalebox{0.8}{
            \setlength\tabcolsep{1pt}
            \renewcommand{\arraystretch}{1.1}
\begin{tabular}{llcccccc}
    \hline
    \multicolumn{1}{c|}{\textbf{Before-hint}} & 
    \multicolumn{1}{c|}{\multirow{2}{*}{\textbf{Reflection}}} &
    \multicolumn{3}{c|}{\textbf{SRL Phase}} &
    \multicolumn{3}{c}{\textbf{Crit. Engag.}}
    \\
    \multicolumn{1}{c|}{\textbf{prompt}} & 
    \multicolumn{1}{c|}{}&
    \multicolumn{1}{c|}{\phantom{0}P\phantom{0}} &
    \multicolumn{1}{c|}{\phantom{0}M\phantom{0}} &
    \multicolumn{1}{c|}{\phantom{0}E\phantom{0}} &
    \multicolumn{1}{c|}{\phantom{0}D\phantom{00}} &
    \multicolumn{1}{c|}{\phantom{0}E\phantom{00}} &
    \multicolumn{1}{c}{\phantom{0}A\phantom{0}}
    \\
    \hline
    \multicolumn{1}{c|}{Planning} & 
    \multicolumn{1}{p{0.65\linewidth}|}{\textit{Right now I need to formulate the problem better I think. (A1Q1)}} &
    \multicolumn{1}{c|}{\checkmark{}} &
    \multicolumn{1}{c|}{} &
    \multicolumn{1}{c|}{} &
    \multicolumn{1}{c|}{\checkmark{}} &
    \multicolumn{1}{c|}{} &
    \multicolumn{1}{c}{}
    \\
    \multicolumn{1}{c|}{Planning} & 
    \multicolumn{1}{p{0.65\linewidth}|}{\textit{I'm stuck on figuring out the regex that distinguishes the text in the user-name column. (A1Q3)}} &
    \multicolumn{1}{c|}{} &
    \multicolumn{1}{c|}{} &
    \multicolumn{1}{c|}{\checkmark{}} &
    \multicolumn{1}{c|}{\checkmark{}} &
    \multicolumn{1}{c|}{\checkmark{}} &
    \multicolumn{1}{c}{}
    \\
    \multicolumn{1}{c|}{Monitoring} & 
    \multicolumn{1}{p{0.65\linewidth}|}{\textit{Consolidating excel files into one and merging dataframes (A3Q1)}} &
    \multicolumn{1}{c|}{} &
    \multicolumn{1}{c|}{} &
    \multicolumn{1}{c|}{\checkmark{}} &
    \multicolumn{1}{c|}{\checkmark{}} &
    \multicolumn{1}{c|}{} &
    \multicolumn{1}{c}{}
    \\
    \multicolumn{1}{c|}{Monitoring} & 
    \multicolumn{1}{p{0.65\linewidth}|}{\textit{loading data says load\_ticket\_data is not a dataframe when it is pd.concat (A3Q1)}} &
    \multicolumn{1}{c|}{} &
    \multicolumn{1}{c|}{} &
    \multicolumn{1}{c|}{} &
    \multicolumn{1}{c|}{\checkmark{}} &
    \multicolumn{1}{c|}{} &
    \multicolumn{1}{c}{}
    \\
    \multicolumn{1}{c|}{Evaluation} & 
    \multicolumn{1}{p{0.65\linewidth}|}{\textit{Perhaps I should create a for loop that adds each match to a list? (A1Q1)}} &
    \multicolumn{1}{c|}{} &
    \multicolumn{1}{c|}{} &
    \multicolumn{1}{c|}{} &
    \multicolumn{1}{c|}{\checkmark{}} &
    \multicolumn{1}{c|}{} &
    \multicolumn{1}{c}{\checkmark{}}
    \\
    \multicolumn{1}{c|}{Evaluation} & 
    \multicolumn{1}{p{0.65\linewidth}|}{\textit{Trying to remove the extra data from the pivot table. Or instead of using a pivot table, use groupby functions (A3Q2)}} &
    \multicolumn{1}{c|}{\checkmark{}} &
    \multicolumn{1}{c|}{} &
    \multicolumn{1}{c|}{\checkmark{}} &
    \multicolumn{1}{c|}{\checkmark{}} &
    \multicolumn{1}{c|}{} &
    \multicolumn{1}{c}{}
    \\
    \hline
    
\end{tabular}
        }
    \end{subtable}
    \begin{subtable}{\linewidth}
        \vspace{3mm}
        \centering
        \subcaption{After-hint Reflection condition}
        \label{fig:pre_post_hint_reflections.post}
        \vspace{-1.5mm}
        \scalebox{0.8}{
            \setlength\tabcolsep{1pt}
            \renewcommand{\arraystretch}{1.1}

\begin{tabular}{llcccccc}
    \hline
    \multicolumn{1}{p{0.17\linewidth}|}{\textbf{After-hint}} & 
    \multicolumn{1}{c|}{\multirow{2}{*}{\textbf{Reflection}}} &
    \multicolumn{3}{c|}{\textbf{SRL Phase}} &
    \multicolumn{3}{c}{\textbf{Crit. Engag.}}
    \\
    \multicolumn{1}{c|}{\textbf{prompt}} &    
    \multicolumn{1}{c|}{}&
    \multicolumn{1}{c|}{\phantom{0}P\phantom{0}} &
    \multicolumn{1}{c|}{\phantom{0}M\phantom{0}} &
    \multicolumn{1}{c|}{\phantom{0}E\phantom{0}} &
    \multicolumn{1}{c|}{\phantom{0}D\phantom{00}} &
    \multicolumn{1}{c|}{\phantom{0}E\phantom{00}} &
    \multicolumn{1}{c}{\phantom{0}A\phantom{0}}
    \\
    \hline
    \multicolumn{1}{c|}{Planning} & 
    \multicolumn{1}{p{0.65\linewidth}|}{\textit{I should be trying to use the VERBOSE form, but I cannot get it to work in that format (A1Q3)}} &
    \multicolumn{1}{c|}{} &
    \multicolumn{1}{c|}{} &
    \multicolumn{1}{c|}{} &
    \multicolumn{1}{c|}{\checkmark{}} &
    \multicolumn{1}{c|}{} &
    \multicolumn{1}{c}{}
    \\
    \multicolumn{1}{c|}{Planning} & 
    \multicolumn{1}{p{0.65\linewidth}|}{\textit{hint was not relevant (A3Q1)}} &
    \multicolumn{1}{c|}{} &
    \multicolumn{1}{c|}{} &
    \multicolumn{1}{c|}{} &
    \multicolumn{1}{c|}{} &
    \multicolumn{1}{c|}{} &
    \multicolumn{1}{c}{}
    \\
    \multicolumn{1}{c|}{Monitoring} & 
    \multicolumn{1}{p{0.65\linewidth}|}{\textit{I think my approach is solid; in fact manually checking my results yields the same answer as my function. I'll parse through the data more; maybe there are letter case issues I haven't considered (A3Q4)}} &
    \multicolumn{1}{c|}{\checkmark{}} &
    \multicolumn{1}{c|}{} &
    \multicolumn{1}{c|}{\checkmark{}} &
    \multicolumn{1}{c|}{\checkmark{}} &
    \multicolumn{1}{c|}{} &
    \multicolumn{1}{c}{}
    \\
    \multicolumn{1}{c|}{Monitoring} & 
    \multicolumn{1}{p{0.65\linewidth}|}{\textit{This hint was both helpful and non-helpful -- helpful as in it's correct, but unhelpful in that I already realized that and that is what I'm spinning my wheels trying to do (A4Q4)}} &
    \multicolumn{1}{c|}{} &
    \multicolumn{1}{c|}{} &
    \multicolumn{1}{c|}{} &
    \multicolumn{1}{c|}{} &
    \multicolumn{1}{c|}{} &
    \multicolumn{1}{c}{}
    \\
    \multicolumn{1}{c|}{Evaluation} & 
    \multicolumn{1}{p{0.65\linewidth}|}{\textit{helpful (A4Q3)}}  &
    \multicolumn{1}{c|}{} &
    \multicolumn{1}{c|}{} &
    \multicolumn{1}{c|}{} &
    \multicolumn{1}{c|}{} &
    \multicolumn{1}{c|}{} &
    \multicolumn{1}{c}{}
    \\
    \multicolumn{1}{c|}{Evaluation} & 
    \multicolumn{1}{p{0.65\linewidth}|}{\textit{I don't think IssueTime is the issue I'm having, I'm struggling with the pivot table (A3Q2)}} &
    \multicolumn{1}{c|}{} &
    \multicolumn{1}{c|}{} &
    \multicolumn{1}{c|}{\checkmark{}} &
    \multicolumn{1}{c|}{\checkmark{}} &
    \multicolumn{1}{c|}{} &
    \multicolumn{1}{c}{}
    \\
    \hline
    
\end{tabular}

        }
    \end{subtable}
\end{table}

In contrast, most after-hint reflections (11 out of 18 non-empty responses) focused on evaluating the relevance or usefulness of the hints provided, rather than on their own problem-solving process. Typical comments included "\textit{hint was not relevant}" (and "\textit{[The support system] gave me the same hint twice in a row (...)}" (See Table~\ref{fig:pre_post_hint_reflections.post}).

Quantitative coding of reflection quality, as shown in Figure~\ref{fig:result_impact_reflection.reflection_qualities}, indicated that self-evaluation was the most common SRL phase present in both conditions, while planning and monitoring were less frequently observed. Notably, none of the reflections in either group demonstrated engagement with the monitoring phase. With respect to Critical Engagement codes, the majority of non-empty reflections described the key issues or events students were facing, but fewer than 10\% addressed why those issues were important or what knowledge would be needed to resolve them.

\subsection{RQ2: Impacts of SRL Phase Associated with Prompts} \label{sec:results.SRL_prompts}

\subsubsection{RQ2a-b: Impact on Hint-seeking Experience and Solving Performance} 
\begin{figure}[t]
    \centering
    \begin{subfigure}{\linewidth}
        \centering
        \scalebox{1}{
            \includegraphics[height=0.03\paperheight]{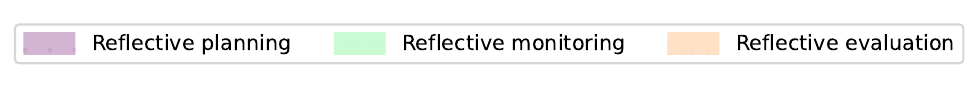}
        }
        \vspace{-2.5mm}
    \end{subfigure}
    \begin{minipage}{\linewidth}
        \begin{subfigure}{0.3\linewidth}
        {
            \centering
            \includegraphics[width=\linewidth]{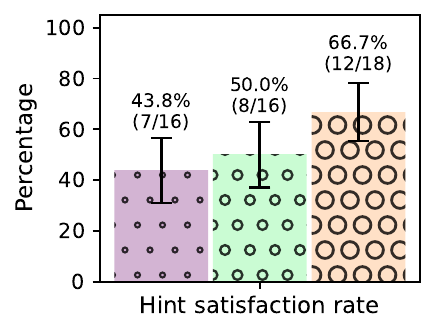}
            \vspace{-6.5mm}        
            \subcaption{Results for RQ2a}
            \label{fig:result_prompts.prehint_reflection_on_hint_satisfaction}
        }
        \end{subfigure}
        \hfill
        \begin{subfigure}{0.3\linewidth}
        {
            \centering
            \includegraphics[width=\linewidth]{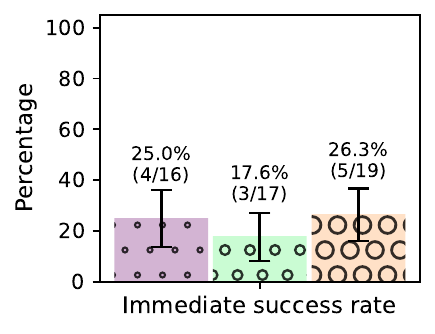}
            \vspace{-6.5mm}        
            \subcaption{Results for RQ2b}
            \label{fig:result_prompts.prehint_reflection_on_immediate_success}
        }
        \end{subfigure}
        \hfill
        \begin{subfigure}{0.3\linewidth}
        {
            \centering
            \includegraphics[width=\linewidth]{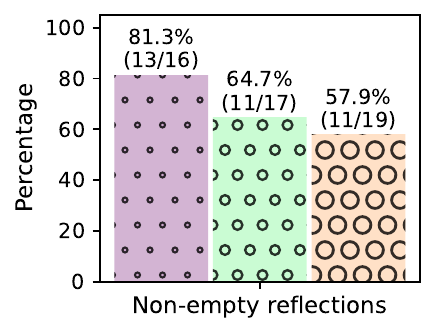}
            \vspace{-6.5mm}        
            \subcaption{Results for RQ2c}
            \label{fig:result_prompts.prehint_reflection_on_non_empty_reflection}
        }
        \end{subfigure}        
    \end{minipage}
    \begin{subfigure}{\linewidth}
        \vspace{2.5mm}
        \centering
        \scalebox{1}{
            \includegraphics[height=0.03\paperheight]{fig/fig4_rq2_legend.pdf}
        }
        \vspace{-2.5mm}
    \end{subfigure}
    \begin{subfigure}{\linewidth}
    {
        \centering
        \includegraphics[width=\linewidth]{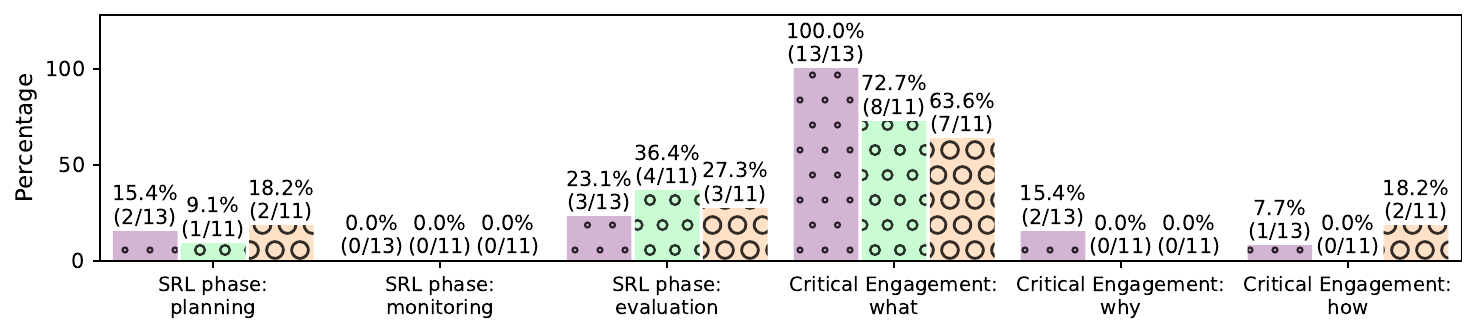}
        \vspace{-6.5mm}
        \subcaption{
            Results for RQ2d
        }
        \label{fig:result_prompts.prehint_reflection_on_reflection_quality}
    }
    \end{subfigure} 
    \vspace{-6.5mm}
    \caption{
        Results for RQ2: Impacts of SRL phase associated with reflection prompts. Similar to Figure~\ref{fig:result_reflection_impacts}, this figure shows the impacts of the SRL phase associated with reflection prompts on: (a) hint satisfaction rate, (b) immediate success rate, (c) non-empty reflection rate, and (d) reflection quality.
    }
    \label{fig:result_prompts}
\end{figure}

Hint-seeking experience was measured through hint satisfaction, and problem-solving performance was measured through immediate success rate (see Section~\ref{sec:analysis_setup.rq2}). As shown in Figure~\ref{fig:result_prompts.prehint_reflection_on_hint_satisfaction}, students are most satisfied with hints following the reflective evaluation prompt ($66.7\%$) and least satisfied with those following the reflective planning prompt ($43.8\%$). The immediate success rate, on the other hand, is highest for the evaluation prompt ($26.3\%$) and lowest for the monitoring prompt ($17.6\%$), as shown in Figure~\ref{fig:result_prompts.prehint_reflection_on_immediate_success}.


\subsubsection{RQ2c-d: Impact on Participation Rate and Reflection Quality} The results for RQ2c and RQ2d are shown in Figures~\ref{fig:result_prompts.prehint_reflection_on_non_empty_reflection} and ~\ref{fig:result_prompts.prehint_reflection_on_reflection_quality}, respectively. We observe varying participation rates, with students writing the most non-empty reflections for the reflective planning prompt ($81.3\%$) and the fewest for the reflective evaluation prompt ($57.9\%$). This pattern also holds in the qualitative coding, in the Critical Engagement code related to the description of key issues or events. Notably, all $13$ non-empty reflections following the reflective planning prompt contain a description of the key issues or events the students were facing (examples are shown in Figure~\ref{fig:pre_post_hint_reflections.pre}).

\subsection{RQ3: Impacts of Prompt Types on Reflection Practices} \label{sec:results.prompt_types}

\subsubsection{RQ3a-b: Impact on Hint-seeking Experience and Solving Performance}

\begin{figure}[]
    \centering
    \begin{subfigure}{\linewidth}
    {
        \centering
        \scalebox{0.94}{
            \includegraphics[width=0.95\linewidth]{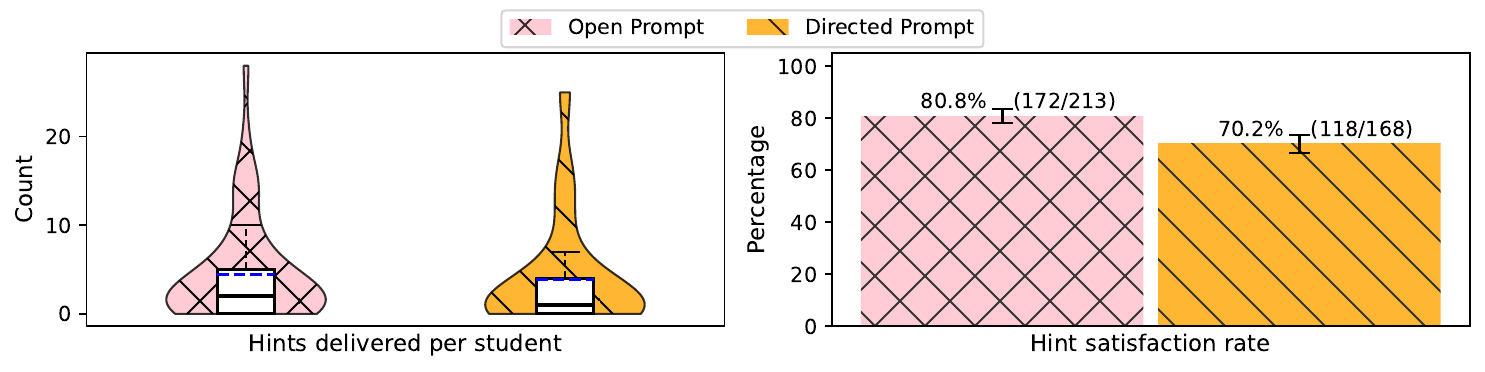}
        }
        \vspace{-2.5mm}
        \subcaption{Results for RQ3a: Similar to Figure~\ref{fig:result_reflection_impacts.help_seeking}, this figure shows the impact of prompt type on students' help-seeking experience.}
        \label{fig:result_prompt_types.a}
    }
    \end{subfigure}
    \begin{subfigure}{\linewidth}
    {
        \vspace{1.5mm}
        \centering
        \scalebox{0.94}{
            \includegraphics[width=0.95\linewidth]{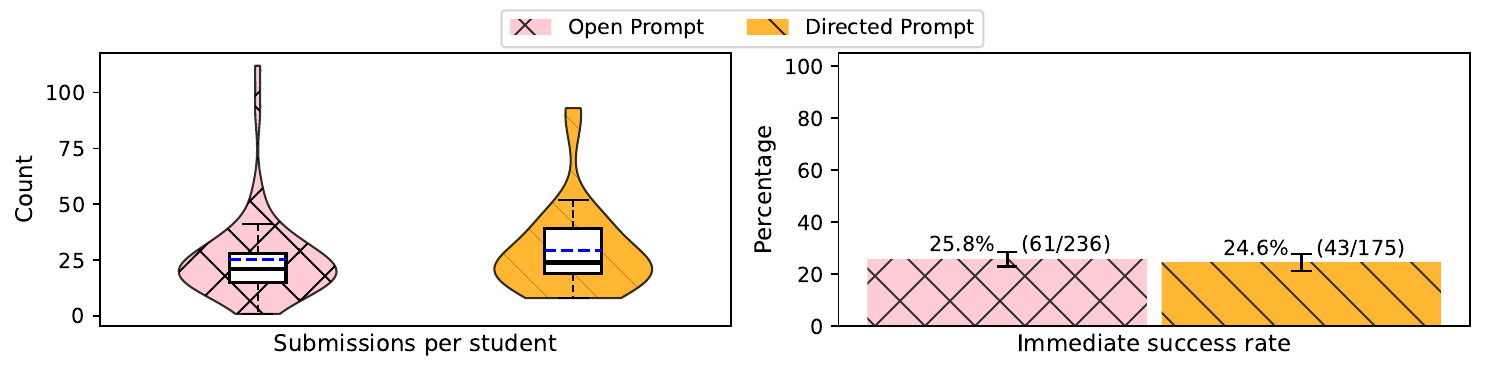}
        }
        \vspace{-2.5mm}
        \subcaption{Results for RQ3b: Similar to Figure~\ref{fig:result_reflection_impacts.solving_performance}, this figure shows the impact of prompt type on students' solving performance.
        }
        \label{fig:result_prompt_types.b}
    }
    \end{subfigure}
    \begin{subfigure}{\linewidth}
    {
        \vspace{1.5mm}
        \centering
        \scalebox{0.94}{
            \includegraphics[width=0.95\linewidth]{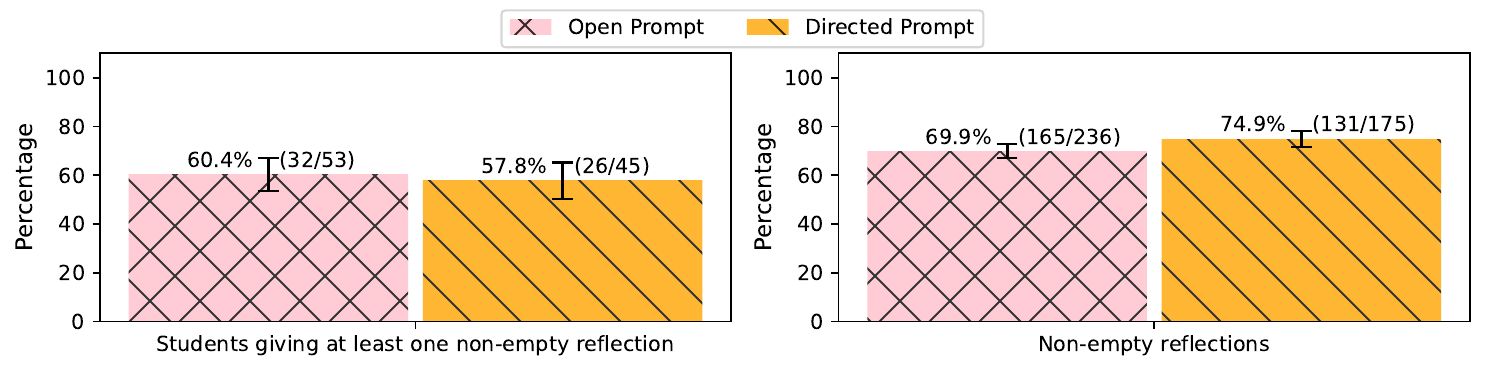}
        }
        \vspace{-2.5mm}
        \subcaption{Results for RQ3c: Similar to Figure~\ref{fig:result_impact_reflection.participation_rate}, this figure shows the impact of prompt type on students' participation in reflection practices.
        }
        \label{fig:result_prompt_types.c}
    }
    \end{subfigure}
    \begin{subfigure}{\linewidth}
    {
        \vspace{1.5mm}
        \centering
        \scalebox{0.94}{
            \includegraphics[width=0.95\linewidth]{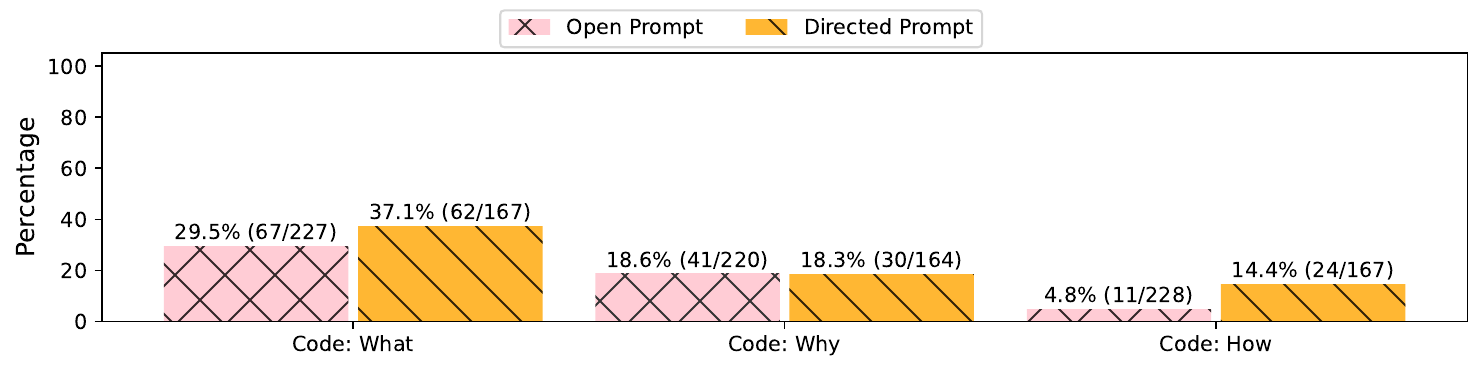}
        }
        \vspace{-2.5mm}
        \subcaption{Results for RQ3d: Similar to Figure~\ref{fig:result_impact_reflection.reflection_qualities}, this figure shows the impacts of prompt type on students' reflection quality. The codes are outlined in Section~\ref{sec:analysis_setup.rq3} and detailed in Table~\ref{tab:reflection_quality_codes}. We note that unresolved cases where the coders could not reach consensus are not included in the analysis: $N_{what} = 6$ (six cases where coders could not agree whether “what” should be applied.), $N_{why} = 16$, $N_{how} = 5$.
        }
        \label{fig:result_prompt_types.d}
    }
    \end{subfigure}
    
    \vspace{-2mm}
    \caption{
        Results for RQ3: Impacts of prompt types on students' hint-seeking experience and solving performance.
    }
    \label{fig:result_prompt_types}
\end{figure}

As shown in Figure~\ref{fig:result_prompt_types.a}, the number of hints per student does not differ much across the two prompt conditions, with the Open Prompt condition slightly higher with 4.5 hints per student compared to the Directed Prompt condition with 3.9. A two-tailed t-test indicates no difference between the two conditions ($t=0.45$, $p=0.650$, $df=96$). In terms of hint satisfaction rate, students in the Open Prompt condition rated hints more highly. A $\chi^2$ test suggests a statistically significant difference in hint rating ($\chi^2=5.71$, $p=0.017$, $df=1$).

Figure~\ref{fig:result_prompt_types.b} illustrates the results for RQ3b. A two-tailed t-test on the number of submissions ($t=-1.02$, $p=0.312$, $df=96$) and a $\chi^2$ test on immediate success rate ($\chi^2=0.09$, $p=0.769$, $df=1$) suggest no difference between the two conditions.


\subsubsection{RQ3c-d: Impact on Participation Rate and Reflection Quality}
Figure~\ref{fig:result_prompt_types.c} shows the results for RQ3c. Two $\chi^2$ tests on the number of students giving non-empty reflections ($\chi^2=0.07$, $p=0.794$, $df=1$) and the number of non-empty reflections ($\chi^2=1.22$, $p=0.270$, $df=1$) suggest no difference between the two conditions.

For RQ3d, Figure~\ref{fig:result_prompt_types.d} summarizes the qualitative coding results for student reflection as responses to the directed and open prompts. Each reflection was evaluated for the presence of three components: ``what'' (describes the current error), ``why'' (explains why the error occurs), and ``how'' (suggests how to resolve the error or its cause). Examples of student reflection to the Open and Directed Prompt types are presented in Tables~\ref{tab:prompt_type_reflections.open} and ~\ref{tab:prompt_type_reflections.directed}, respectively.

\begin{table}[t!]
     \caption{
        Examples of reflections written by students in the Open and Directed Prompt conditions (in Trial 2) with annotation of theme Critical Engagement. The expert annotation process is described in Section~\ref{sec:analysis_setup.rq3} and the complete list of codes is shown in Table~\ref{tab:reflection_quality_codes}. The assignment (A) and question (Q) number followed every response. 
    }
    \label{tab:prompt_type_reflections}
    \vspace{-1.5mm}
    \centering
    \begin{subtable}{\linewidth}
        \centering
        \subcaption{Open Prompt condition}
        \label{tab:prompt_type_reflections.open}
        \vspace{-1mm}
        \scalebox{0.8}{
            \setlength\tabcolsep{3pt}
            \renewcommand{\arraystretch}{1.25}
\begin{tabular}{lccc}
    \hline
    \multicolumn{1}{c|}{\multirow{2}{*}{\textbf{Reflection}}} &
    \multicolumn{3}{c}{\textbf{Code}}
    \\
    \multicolumn{1}{c|}{}&
    \multicolumn{1}{c|}{What} &
    \multicolumn{1}{c|}{Why} &
    \multicolumn{1}{c}{How}
    \\
    \hline
    \multicolumn{1}{p{0.95\linewidth}|}{\textit{[Student' regex] pattern needs work? (A1Q3)}} &
    \multicolumn{1}{c|}{} &
    \multicolumn{1}{c|}{} &
    \multicolumn{1}{c}{}
    \\
    \multicolumn{1}{p{0.95\linewidth}|}{\textit{My dictionary might not be saved correctly, as I'm getting an assert error that I'm not returning a dict	(A2Q1)}} &
    \multicolumn{1}{c|}{\checkmark{}} &
    \multicolumn{1}{c|}{\checkmark{}} &
    \multicolumn{1}{c}{}
    \\
    \multicolumn{1}{p{0.95\linewidth}|}{\textit{I still cannot find the column names you're looking for. I just don't know what they are. What are they? I won't be able to find out otherwise (A2Q3)}} &
    \multicolumn{1}{c|}{} &
    \multicolumn{1}{c|}{} &
    \multicolumn{1}{c}{}
    \\
    \multicolumn{1}{p{0.95\linewidth}|}{\textit{could you take a look at the codes for question 1? I am having a hard time to match out the total number of rows (A3Q1).}} &
    \multicolumn{1}{c|}{\checkmark{}} &
    \multicolumn{1}{c|}{} &
    \multicolumn{1}{c}{}
    \\
    \multicolumn{1}{p{0.95\linewidth}|}{\textit{What could be wrong with my percentage delays? United airlines isn't the bottom one (A4Q3)}} &
    \multicolumn{1}{c|}{\checkmark{}} &
    \multicolumn{1}{c|}{\checkmark{}} &
    \multicolumn{1}{c}{}
    \\
    \multicolumn{1}{p{0.95\linewidth}|}{\textit{What's wrong? (A4Q3)}} &
    \multicolumn{1}{c|}{} &
    \multicolumn{1}{c|}{} &
    \multicolumn{1}{c}{}
    \\
    \hline
    
\end{tabular}
        }
    \end{subtable}
    \begin{subtable}{\linewidth}
        \vspace{3mm}
        \centering
        \subcaption{Directed Prompt condition}
        \label{tab:prompt_type_reflections.directed}
        \vspace{-1mm}
        \scalebox{0.8}{
            \setlength\tabcolsep{3pt}
            \renewcommand{\arraystretch}{1.25}
\begin{tabular}{lccc}
    \hline
    \multicolumn{1}{c|}{\multirow{2}{*}{\textbf{Reflection}}} &
    \multicolumn{3}{c}{\textbf{Code}}
    \\
    \multicolumn{1}{c|}{}&
    \multicolumn{1}{c|}{What} &
    \multicolumn{1}{c|}{Why} &
    \multicolumn{1}{c}{How}
    \\
    \hline
    \multicolumn{1}{p{0.95\linewidth}|}{\textit{I received the following error: You have failed this test due to an error. The traceback has been removed because it may contain hidden tests. This is the exception that was thrown: AssertionError: Sorry, this item should be in the log results, check your formating. Does this mean I have stored something in the wrong variable? (A1Q3)}} &
    \multicolumn{1}{c|}{} &
    \multicolumn{1}{c|}{\checkmark{}} &
    \multicolumn{1}{c}{}
    \\
    \multicolumn{1}{p{0.95\linewidth}|}{\textit{is my answer correct\, just with a negative instead of positive correlation? i'm pretty sure my code is right! (A2Q3)}} &
    \multicolumn{1}{c|}{} &
    \multicolumn{1}{c|}{\checkmark{}} &
    \multicolumn{1}{c}{}
    \\
    \multicolumn{1}{p{0.95\linewidth}|}{\textit{I believe my code is correct, It has the right shape and when i inspect it through snapshots it seems to be right. does it look ok (A3Q1)}} &
    \multicolumn{1}{c|}{} &
    \multicolumn{1}{c|}{} &
    \multicolumn{1}{c}{\checkmark{}}
    \\
    \multicolumn{1}{p{0.95\linewidth}|}{\textit{I have all the correct counts except vanity\_count. I believe that the error comes in when I calculate total\_plates. I don't think I should have the .notna() function, but the code doesn't run without it (A3Q4)}} &
    \multicolumn{1}{c|}{\checkmark{}} &
    \multicolumn{1}{c|}{\checkmark{}} &
    \multicolumn{1}{c}{\checkmark{}}
    \\
    \multicolumn{1}{p{0.95\linewidth}|}{\textit{i am lost, i am getting an error without a specific reason (A4Q3)}} &
    \multicolumn{1}{c|}{} &
    \multicolumn{1}{c|}{} &
    \multicolumn{1}{c}{}
    \\
    \multicolumn{1}{p{0.95\linewidth}|}{\textit{One bug that I can easily identify is the shape, but I can't tell which row is extraneous, considering I'm expecting 11,5 and getting 12,5. My biggest question is why I'm only getting NaN values in my resulting dataframe. (A4Q4)}} &
    \multicolumn{1}{c|}{\checkmark{}} &
    \multicolumn{1}{c|}{\checkmark{}} &
    \multicolumn{1}{c}{}
    \\
    \hline
    
\end{tabular}
        }
    \end{subtable}
\end{table}

To examine differences in the frequency of each qualitative code between the prompt types, $\chi^2$ tests were conducted. For the ``what'' code ($\chi^2=2.20$, $p=0.138$, $df=1$) and the ``why'' code ($\chi^2=0.00$, $p=1.000$, $df=1$), no significant difference was observed. However, for the ``how'' code, a significant difference was found between conditions ($\chi^2=9.73$, $p=0.002$, $df=1$).


\section{Discussion} \label{sec:discussion}
In this section, we discuss the insights and implications for AI design and development in educational settings. First, we will go through the primary insights and implications based on our overall findings, and then zoom down on insights and implications from answers to specific RQs.  

\subsection{Tension between Pedagogical Values and AI Optimization Methods}

\subsubsection{Summary of the Findings}
Across all three research questions, our findings reveal a tension between student satisfaction with AI-generated hints and the quality of their reflection. Conditions that yielded higher satisfaction consistently corresponded to lower-quality reflection practices, as measured through hint satisfaction rates, reflection depth, and participation metrics.
In RQ1, students who were prompted to reflect after receiving hints (after-hint) demonstrated higher (though not statistically significant) satisfaction compared to the before-hint group. However, the before-hint group exhibited significantly more higher participation rate of reflection and demonstrated more advanced reflection levels. 
For RQ2, the reflective evaluation prompt resulted in the highest hint satisfaction but the lowest rates of both meaningful and advanced reflections. Conversely, the reflective planning prompt led to the lowest satisfaction but the highest rates of participation and responses with more sophisticated reflection components.
In RQ3, students in the open prompt condition reported significantly higher hint satisfaction rates compared to the directed prompt condition. While differences in participation rate were not statistically significant, students in the directed prompt type were statistically significantly more likely to produce constructive reflection responses that included deeper reflection components (``how''), despite reporting significantly lower satisfaction with the hints. 

On the other hand, no clear patterns were found between the immediate success rate and either hint satisfaction or reflection quality.

\subsubsection{Theoretical Interpretation}
The inverse relationship between quality reflection and satisfaction with AI-generated hints echoes established research on the dynamics between effortful learning and satisfaction of students. 
Reflection, as a component of SRL, demands substantial cognitive load~\citep{de2020effort,seufert2024}.
Previous research demonstrates that student satisfaction generally decreases as mental effort increases, especially when cognitive load exceeds manageable levels.
\citet{bradford2011cognitive} found that cognitive load accounted for about 25\% of the variance in satisfaction in online learning. Similarly, \citet{ma2025investigation} found that cognitive load had a statistically significant negative effect on English-as-a-foreign language (EFL) students' satisfaction with MOOCs. A decrease in satisfaction and willingness on learning tasks was likely to develop into disengagement with learning activities.

Existing SRL theoretical frameworks focusing on affective aspects could provide important foundations to balance effortful learning with AI assistance for students to continue engaging meaningfully with learning activities. Efklides' Metacognitive and Affective model of SRL (MASRL)~\citep{efklides2011} emphasizes the interactions between metacognition, motivation, and affect across two levels of functioning: the Person level (stable characteristics) and the Task X Person level (dynamic task-specific interactions). Boekaerts' Dual Processing model~\citep{boekaerts2011} describes how students navigate between growth pathways and well-being pathways based on their appraisals of learning situations, including their emotional responses. Integrating those SRL frameworks into educational research in technology-rich educational contexts would add value by advancing structured understanding of ways to embed AI assistance for more benefits and less harm on student SRL.

\subsubsection{Systemic Implication}
Our findings, combined with educational theories, call for systemic changes in how AI models are trained, evaluated, and aligned for educational use. Current AI models are primarily trained to optimize their behavior in accordance with user preferences, as reflected in reinforcement learning from human feedback~\citep{ouyang2022training}. However, user preferences in education often favor convenience and ease over pedagogical value. As a result, AI systems may perform tasks that allow learners to bypass cognitive effort, even when effort is essential for skill development. 

This finding highlights the importance of educational AI system designs that incorporate the concept of desirable difficulties rather than solely pursuing user satisfaction maximization~\citep{bjork1994memory,bjork2011making}. Prioritizing satisfaction alone in AI optimization risks diminishing opportunities for students to `struggle' productively and develop SRL skills (e.g., reflection) and other fundamental learning competencies such as critical thinking. Many studies have already shown that most learning occurs as students are challenged; they learn as they productively struggle~\citep{warshauer2015productive, warshauer2015strategies}, work on tasks with desirable difficulty~\citep{bjork2011making} that students independently cannot resolve but can with others' assistance~\citep{vygotsky1978mind}. It is a timely issue to move to a focus on inferring and providing the zone of proximal difficulty as industry leaders, researchers, and practitioners examine, adopt, and create AI-powered educational tools. While the majority of AI-powered tools are developed to effectively assist students, our study findings draw our attention to AI development to sometimes have students struggle. 

Our study findings also call for the transformation of the learning evaluation frameworks. Our results show negligible effects on immediate success rates, a metric used for problem-solving performance. Under the current practices of summative evaluation of short-term performance, the trade-off between learning opportunity and satisfaction would not be easily detected. Thus, it is important for AI researchers in educational settings to adopt diverse metrics beyond performance to evaluate the impact of AI use on learning~\citep{bjork1994memory,bjork2011making}.

In our study, we mitigated this problem through context engineering: explicitly signaling the educational intent in the AI-prompt and designing non-conversational user interactions to limit pedagogically irrelevant information 
While effective to some extent, such design choices also limit interaction depth. A more scalable solution could involve pre-training foundational models on pedagogically aligned data. This would allow for more robust multi-turn dialogue with learners while maintaining a focus on learning goals, not just user satisfaction.

Foundation models are typically trained on large, diverse text datasets and aligned with human preferences using techniques such as Direct Preference Optimization (DPO) \citep{rafailov2023direct} and Reinforcement Learning through Human Feedback (RLHF)~\citep{ouyang2022training}. Multiple methods exist for preference optimization, including ordinal ranking of responses~\citep{christiano2017deep}, binary feedback through user voting mechanisms, as employed in this study and typical in conversational systems~\citep{stiennon2020learning}, and AI-based response scoring for safety guardrails \citep{bai2022constitutional}. However, none of these methods address the subjective qualities underlying preference definitions. Current preference alignment approaches generally rely on end-user opinions. While correctness is often used as a preference criterion, pedagogical theories, such as theories of metacognition~\citep{efklides2011}, SRL~\citep{zimmerman2003motivating}, and productive struggle~\citep{bjork1994memory}, are not integrated into model alignment processes. Integrating these theoretical constructs into the model alignment process could lead to AI systems better suited for long-term learning outcomes rather than short-term satisfaction.

In short, our findings reveal a fundamental tension between AI optimization objectives and pedagogical effectiveness. It demands reconsideration of how AI systems are designed and trained for educational contexts. The evidence suggests that prioritizing immediate user satisfaction may undermine the deeper learning processes that educational AI systems should support.

\subsubsection{Limitations and Future Work}
Several limitations must be noted.
Our study was conducted in a fully online programming course with adult learners. Future research should examine how these dynamics play out with K-12 students, who are still developing SRL and higher-order thinking skills. 
We disabled conversational interaction between students and the AI to avoid gaming behaviors. This design choice limited our ability to explore how dialogue types might affect student satisfaction-reflection dynamics, such as the impacts of AI response types on students' metacognition and cognition activities. 
Our study did not incorporate human-in-the-loop designs. Exploring how human instructors and tutors might ease the discovered tension and complement our AI-powered tool remains a critical area for future work.

\subsection{The Impact of Prompt Types with AI-Generated Hints}
\subsubsection{Summary of the Findings}
In Trial 1, addressing RQ1a-b, students in both experimental groups (prompts \emph{before} and \emph{after} hints) tended to request fewer AI-generated hints than the Control group (no prompt), although this difference was not statistically significant. However, the experimental groups exhibited narrower interquartile ranges in the number of hint requests, suggesting more consistent behavior. No significant differences were found in immediate success rates or assignment submission counts across conditions.
Between \emph{before}- and \emph{after}-hint conditions (RQ1c-d), prompting students to reflect \emph{before} receiving AI-generated hints led to higher participation with reflection practices, more constructive reflection responses, and lower satisfaction with the hints. In contrast, students prompted \emph{after} receiving hints often skipped the prompt or gave brief, superficial comments (e.g., ``helpful'').

For RQ2, although no statistical tests were performed due to a small data size, trends showed that for the \emph{before}-hint condition, the reflective \emph{evaluation} prompt was associated with the highest satisfaction with hints. However, it led to the lowest participation rate in reflection, while the \emph{planning} prompt resulted in the highest participation.

Regarding RQ3, \emph{open} prompt group students produced less constructive reflection responses compared to \emph{directed} prompts; their responses often had a description of ``what'' the current error was, but fewer provided analysis of ``why'' the error occurred or ``how'' to address it. 

\subsubsection{Theoretical Interpretation}
In our study, experimental groups compared to Control (RQ1) and directed prompt groups compared to open prompts (RQ3) tended to ask for a lower number of hints. Yet, the difference was not significant. 
Previous research has reported mixed effects of reflective activities, including self-evaluation, on students’ help-seeking frequency ~\citep{roll2011improving, daley2016beyond}. The narrower interquartile ranges of both before- and after-hint groups might be worthy of further investigation. One hypothesis to explain the results would be that confounding variables, such as relevant skill competencies or prior knowledge, could have interfered with the impact of the conditions on the number of hint requests. The impact of the conditions might have still decreased the need for more hints of struggling students, resulting the narrower ranges compared to the Control group.

Impacts on reflection quality display clearer patterns throughout trials. \emph{After}-hint reflection might have been not as effective as \emph{before}-hint reflection (RQ1). This is because exposure to hints prior to reflection might have led students to focus on using the hints to find the correct answer, rather than pausing to analyze the problem in their own words~\citep{edwards2017reflecting, baars2020effort,moon2004reflection}. In addition, some students might have thought that their reflection responses would be used as input to generate more personalized AI-generated hints, even though our instructions clearly conveyed that their response was not used for hint generation. While there is not much research on direct comparison between planning and evaluation in the context of reflection prompting, one potential explanation would be that \emph{planning} might have felt more actionable and aligned with a type of assistance for which students sought. While \emph{planning} and \emph{evaluation} are not two clearly separate phases, \emph{planning} prompts might have been more associated with progress and pathway forward (e.g., ``(...) what are the steps you think must be followed in order to answer this question (...),'' similar with what they expected from hints. This pattern is consistent with previous research, which has shown that students are more likely to engage in lower-level reflections, such as description, than in higher-level reflections including integration and interpretation~\citep{Bain1999-ji,Bain2002-qc,ryan2014reflection}. On the other hand, it might have felt more challenging for students to \emph{evaluate} their problems and approaches at which they were stuck  (``(...) is there an alternative approach which you can try to solve the step of the question (...)''). Additionally, students are shown to find it easier to reflect on current challenges than to evaluate alternatives, which can be more cognitively demanding~\citep{lin2022metacognitive}. Finally, the directed prompt with more specific guidance (RQ3) could have been more effective at leading students toward quality reflection, especially to students who were not familiar with reflections and more likely to be tempted to metacognitive laziness to avoid frustration. \citet{zhang2024students} found that prompts providing more instruction improved the performance and response quality of students who were less familiar with the self-explanation strategy. In contrast, students who were already familiar with the strategy showed no difference in the benefits of more instructional versus open prompts.

\subsubsection{Practical Implications}

These findings underscore the importance of well-designed reflection interventions, especially when students are experiencing high cognitive load and AI-powered tools are a click away. Well-designed reflection prompting could guide students to deliberately pause to engage with reflections, an effortful and beneficial learning process, instead of hasty application of hints. Furthermore, students could also practice and integrate higher-level reflection (analysis of ``why'' they are stuck or ``how'' to integrate hints to resolve their problems) to their real-world learning processes~\citep{Bain1999-ji,Bain2002-qc,ryan2014reflection}. Specifically, this study informs the decision of when to provide hints, which SRL phases to target, and how much guidance to provide. 

\subsubsection{Limitations and Future Work}
This study did not measure the long-term development of students' reflection skills or performance. Future research should adopt a longitudinal design to assess how reflection practices combined with AI-generated hints influence reflective growth and learning outcomes.
Due to our sample size and the measures collected, definitive conclusions about when and how reflection practices affect AI support usage and learning gains remain elusive. Future studies with larger, more diverse cohorts should investigate whether reflection prompts can consistently reduce students' dependence on AI assistance, elevate satisfaction through metacognitively informed hinting, and foster equitable SRL across ability levels.
Finally, interactions between reflection prompting and student competency warrant closer examination. This understanding could clarify under which conditions and for which students reflection practices meaningfully impact help-seeking behavior.


\section{Conclusions}

This study advances understanding of how reflective scaffolding can be integrated with AI-generated hints to shape student learning in programming courses. Our findings reveal that while reflection before receiving hints and the use of planning-focused or directed prompts foster higher-quality reflection, these approaches are often associated with lower student satisfaction. Moreover, a consistent negative relationship emerged between satisfaction and reflection quality, and immediate problem-solving performance was largely unaffected by reflection interventions. These results highlight a fundamental tension between optimizing for user satisfaction and promoting deeper, self-regulated learning. Moving forward, educational AI systems should be intentionally designed to prioritize pedagogical value and support the development of students’ critical thinking and independent learning skills, rather than focusing solely on satisfaction metrics. This work provides insights for researchers, educators, instructional designers, and AI developers seeking to align AI-powered support with long-term educational goals.




 \bibliographystyle{elsarticle-harv} 
 \bibliography{main}






\end{document}